\documentclass[amssymb,aps]{revtex4-2}

\usepackage[utf8]{inputenc}
\usepackage[english]{babel}
\usepackage[T1]{fontenc}
\usepackage{amsmath}
\usepackage{amssymb}
\usepackage{mathrsfs} 
\usepackage{graphicx}
\usepackage{float}
\usepackage{tikz}
\usepackage{pgfplots}
\pgfplotsset{compat=1.18} 
\usetikzlibrary{decorations.pathmorphing, arrows.meta, positioning}

\usepackage[colorlinks=true, linkcolor=blue, citecolor=blue, urlcolor=blue]{hyperref}

\begin{document}

\title{Universal and rigorous asymptotic convergence of standard macroscopic quantum electrodynamics to the modified Langevin noise formalism}

\author{Alessandro Ciattoni}
\email{alessandro.ciattoni@spin.cnr.it}
\homepage{https://orcid.org/0000-0002-1636-8681}
\affiliation{Consiglio Nazionale delle Ricerche (CNR), SPIN Institute, c/o Dip.to di Scienze Fisiche e Chimiche, Via Vetoio, 67100 Coppito (L'Aquila), Italy}

\begin{abstract}
The standard Langevin noise formalism (LNF) is a well-established approach for macroscopic quantum electrodynamics in unbounded lossy media, and accordingly, its application to finite objects requires endowing the surrounding vacuum with a fictitious absorption that is taken to asymptotically vanish. Existing justifications for such a singular limiting procedure have been mainly restricted to phenomena governed by vacuum fluctuations and medium-induced interactions (Casimir-Polder forces, thermal radiation, quantum emitter dynamics, etc.), thereby strictly excluding scenarios that involve the scattering of external quantum radiation. Conversely, the modified Langevin noise formalism (MLNF) has recently been established as the complete canonical quantum theory for arbitrary macroscopic systems and electromagnetic processes, entirely bypassing the need for limiting procedures, treating incident quantum radiation fields as independent degrees of freedom, and exactly reducing to the LNF for unbounded media. In this work, we demonstrate that the LNF limiting procedure is universally valid by showing that it recovers any physical prediction of the MLNF across all regimes, including arbitrary quantum scattering processes. Specifically, we introduce a suitable mapping associating any MLNF state and observable with a corresponding entity within the LNF augmented by fictitious vacuum absorption. We then rigorously prove that, as the fictitious absorption tends to zero, every LNF matrix element constructed via this mapping exactly converges to its MLNF original, so that the limiting procedure establishes the asymptotic convergence of the LNF to the MLNF.
\end{abstract}

\maketitle

\section{Introduction}

Macroscopic quantum electrodynamics, also referred to as the Langevin noise formalism (LNF), has emerged over the past few decades as an indispensable theoretical framework for describing the interaction between quantized electromagnetic fields and dispersive, absorptive, and inhomogeneous media \cite{Scheel2008, Buhmann2012}. Building upon the foundational reservoir quantization models for lossy dielectrics \cite{Huttner1992, Gruner1995}, the three-dimensional LNF was formulated as a phenomenological approach by Gruner and Welsch \cite{Gruner1996} and further developed throughout the late 1990s \cite{Dung1998, Scheel1998}. Guided by the fluctuation-dissipation theorem, this early approach phenomenologically established the theory by adopting standard free-space commutation relations for the macroscopic fields. This formalism successfully evaluated a vast array of fluctuation-induced phenomena, including Casimir-Polder forces \cite{Buhmann2004, Buhmann_Book1}, quantum friction \cite{Buhmann_Book2}, and spontaneous emission rates \cite{Dung2000}. A rigorous formal foundation for this approach was established by Philbin \cite{Philbin2010}, who demonstrated that for strictly unbounded lossy media, the phenomenological LNF exactly coincides with the second-quantized form of the canonical quantization of macroscopic electromagnetism (CQME).

The structural reliance of the LNF on an infinitely extended dissipative bath poses a twofold critical challenge when attempting to model finite-sized objects isolated in vacuum. First, because the LNF assumes that all fields decay exponentially at spatial infinity, it poses formal challenges when modeling the reality of a lossless vacuum. Consequently, employing the LNF for widely studied effects—such as fluctuation-induced phenomena or quantum emitter dynamics—demands a strict formal justification. Second, the pervasive lossy environment exponentially damps incoming photons, structurally omitting free-space radiation degrees of freedom. While this absence is harmless for the aforementioned phenomena, as they do not involve incident radiation, it makes the formal description of quantum scattering computationally and conceptually intricate within the LNF. To apply the LNF to finite objects, it became standard operational practice \cite{Gruner1996, Dung1998, Scheel2008, Buhmann_Book1} to mimic the vacuum with a fictitious absorbing background $\varepsilon = 1 + i\eta$ and evaluate the limit $\eta \to 0^+$ at the end of the calculations, a procedure whose formal justification has become a subject of recent fundamental interest. While the consistency of the limiting procedure for processes strictly devoid of incident radiation has been validated—for instance, using quasinormal modes \cite{Franke2020}—scenarios involving free-space radiation have been successfully addressed only for pure empty space, where the limit retrieves standard vacuum quantum electrodynamics \cite{Hanson2021}. To the best of our knowledge, the limiting procedure has been applied to quantum scattering only in an early one-dimensional derivation \cite{Gruner1996_InputOutput}, where it served purely as an operational tool without an explicit formal justification. Consequently, it remains an open question whether such a singular limiting procedure can be theoretically justified for any arbitrary macroscopic electrodynamical process, thereby enabling the LNF to also accurately describe three-dimensional quantum scattering by a finite, dispersive, and lossy object.

To completely bypass singular limiting procedures, the incident radiation field must be natively incorporated as an independent degree of freedom, a principle that forms the very foundation of the MLNF. Initially introduced for one-dimensional planar structures \cite{DiStef2001} and subsequently extended to non-magnetic systems \cite{Dreze2017_QuaPol, Dorie2019, Naaaa2023}, the framework has recently been generalized in Ref.~\cite{Ciatt2024} to arbitrary magneto-dielectric objects whose electric permittivity $\varepsilon_\omega(\mathbf{r})$ and magnetic permeability $\mu_\omega(\mathbf{r})$ are subject to no restrictions other than Kramers-Kronig relations; hereafter, we shall refer to this most comprehensive variant simply as the MLNF. A definitive formal basis for this approach was ultimately established in Ref.~\cite{Ciatt2026_DirDer}, where the MLNF was rigorously shown to be the exact second-quantized form of Philbin's CQME \cite{Philbin2010} for finite objects in vacuum. This was achieved by providing the exact analytical expressions for the scattering, electric, and magnetic polariton operators in terms of the canonical CQME field operators, mathematically proving their bosonic nature, and demonstrating that they exactly diagonalize the CQME Hamiltonian. Crucially, that derivation resolved the apparent ``second quantization paradox'': it demonstrated that the original reduction of the CQME to the standard LNF in Ref.~\cite{Philbin2010}  was a direct consequence of adopting a restricted Fano-diagonalization ansatz, which is naturally suited for strictly unbounded dissipative media. Consequently, the MLNF provides the exact, limit-free quantum framework for any arbitrary bounded or unbounded, lossy or transparent, magneto-dielectric system. By construction, it natively encompasses pure vacuum QED, the Glauber-Lewenstein quantization for lossless media \cite{Glauber1991}, and the standard LNF for unbounded lossy media as immediate topological limits. Owing to its inherent capability to explicitly handle quantized scattering modes, the MLNF has recently been exploited to model the non-equilibrium interaction of quantum emitters with dispersive environments \cite{Miano2025_QuaEmi, Miano2025_SpeDen, Miano2026}, to develop a general theoretical platform for macroscopic quantum optical scattering \cite{Ciatt2025_QuaSca, Ciatb2025_ClaTun, Ciattoni2026_FarFie}, and to predict the fluctuation-induced mechanical interactions exerted by arbitrary incoming quantum radiation upon finite absorbing objects at finite temperatures \cite{Ciattoni2026_QuaOpt}.

In this work, we demonstrate the universal validity of the LNF limiting procedure. Specifically, we rigorously prove that, under the asymptotic limit $\eta \to 0^+$, the LNF exactly recovers all physical predictions of the MLNF, including arbitrary quantum scattering processes. Our derivation is grounded in the MLNF framework of Ref.~\cite{Ciatt2026_DirDer} and relies on the identification, within the LNF with fictitious absorption, of specific operators that formally assume the role of the MLNF polariton operators as $\eta$ asymptotically approaches zero.  This identification is achieved through a rigorous analysis of the LNF electric field operator and its sources, demonstrating that it formally reproduces the MLNF field as $\eta \to 0^+$: the contribution from the object's volume directly yields the mapped material polariton operators, whereas the part originating from the fictitious vacuum provides the mapped scattering polariton operators. Exploiting this formal correspondence, we establish a systematic mapping that unambiguously associates any given physical state and operator within the exact MLNF with a counterpart in the LNF with fictitious absorption. We then rigorously prove that, in the limit $\eta \to 0^+$, every matrix element evaluated within this LNF setup exactly converges to the true physical matrix element of the MLNF. Consequently, we conclude that any arbitrary problem in macroscopic quantum electrodynamics can be validly addressed using the LNF equipped with limiting procedures.

The paper is structured as follows. In Sec.~II, we briefly review the MLNF and rigorously demonstrate that the standard LNF is exactly recovered as the limiting case for an unbounded dissipative medium. In Sec.~III, we introduce the ${\rm LNF}^{(\eta)}$ framework endowed with a fictitious vacuum absorption and formalize the mathematical limiting procedure. Section~IV is devoted to mapping the MLNF scattering sector into the ${\rm LNF}^{(\eta)}$ by identifying the specific operators that serve as the formal counterparts to the MLNF scattering polaritons. In Sec.~V, we establish the fundamental mapping between the states and operators of the two frameworks and mathematically prove the universal asymptotic convergence of the associated matrix elements, illustrating the procedure with an explicit physical example. In Sec.~VI, we discuss the physical and methodological implications of our results, situating our derivation within the context of existing literature. Finally, Sec.~VII outlines our conclusions. The rigorous mathematical proofs supporting our analytical derivation are deferred to the Appendices.

\section{Modified and standard Langevin noise formalisms}
In this section, we first briefly review the MLNF, which provides the overarching quantum description of arbitrary macroscopic magneto-dielectric systems. Subsequently, we revisit the LNF by providing the hitherto missing rigorous demonstration that it is exactly recovered as a limiting case of the MLNF when considering the specific geometric limit of a strictly unbounded dissipative medium.

\subsection{MLNF} 
Adopting the formalism and notation established in Ref.~\cite{Ciatt2026_DirDer}, we consider a finite, lossy magneto-dielectric body occupying a spatial region $V$ and embedded in infinite vacuum. Assuming a harmonic time dependence $e^{-i \omega t}$ ($\omega >0$), its macroscopic, inhomogeneous, and isotropic optical response is governed by a relative dielectric permittivity and magnetic permeability that we globally define over the entire space as:
\begin{align} \label{eps_mu}
\varepsilon _\omega  \left( {\bf{r}} \right) =& \left\{ {\begin{array}{*{20}l}
   {\varepsilon _\omega ^V \left( {\bf{r}} \right),} & {{\bf{r}} \in V}  \\
   {1,} & {{\bf{r}} \notin V}  \\
\end{array} } \right. &
\mu _\omega  \left( {\bf{r}} \right) =& \left\{ {\begin{array}{*{20}l}
   {\mu _\omega ^V \left( {\bf{r}} \right),} & {{\bf{r}} \in V}  \\
   {1,} & {{\bf{r}} \notin V.}  \\
\end{array}} \right.
\end{align}
To ensure strict causality, the object's internal profiles $\varepsilon _\omega ^{V} \left( {\bf{r}} \right)$ and $\mu _\omega ^{V} \left( {\bf{r}} \right)$ are required to be holomorphic functions of the complex frequency $\omega$ in the upper half-plane ${\mathop{\rm Im}\nolimits} \left( \omega  \right) > 0$. 

The MLNF describes the field-matter system through three classes of bosonic operators: the scattering ($s$) polariton operators ${\bf{\hat g}}_{\omega s} \left( {\bf{n}} \right)$, which depend on a propagation unit vector $\bf{n}$ and satisfy the transversality condition ${{\bf{n}} \cdot {\bf{\hat g}}_{\omega s} \left( {\bf{n}} \right) = 0}$, and the electric ($e$) and magnetic ($m$) polariton operators ${\bf{\hat f}}_{\omega e } \left( {\bf{r}} \right)$ and ${\bf{\hat f}}_{\omega m } \left( {\bf{r}} \right)$, strictly confined within the object's volume (${{\bf{r}} \in V}$) and vanish identically in the surrounding vacuum. The polariton operators obey the  bosonic commutation relations:
\begin{align} \label{MLNF_Com_Rel}
\left[ {{\bf{\hat g}}_{\omega s} \left( {\bf{n}} \right),{\bf{\hat g}}_{\omega 's}^\dag  \left( {{\bf{n}}'} \right)} \right] &= \delta \left( {\omega  - \omega '} \right)\delta \left( {o_{\bf{n}}  - o_{{\bf{n}}'} } \right){\cal I}_{\bf{n}}, & \left[ {{\bf{\hat f}}_{\omega \nu } \left( {\bf{r}} \right),{\bf{\hat f}}_{\omega '\nu '}^\dag  \left( {{\bf{r}}'} \right)} \right] &= \delta \left( {\omega  - \omega '} \right)\delta _{\nu \nu '} \delta \left( {{\bf{r}} - {\bf{r}}'} \right){\cal I},
\end{align}
with all remaining cross-commutators equating to zero. Here, ${\cal I}$ is the three-dimensional dyadic identity, while ${\cal I}_{\bf{n}} = {\cal I} - {\bf{n}}{\bf{n}}$ denotes the dyadic projector onto the transverse plane perpendicular to the direction ${\bf{n}} = \sin \theta _{\bf{n}} \left( {\cos \varphi _{\bf{n}} {\bf{u}}_x  + \sin \varphi _{\bf{n}} {\bf{u}}_y } \right) + \cos \theta _{\bf{n}} {\bf{u}}_z$. Additionally, $\delta \left( {o_{\bf{n}}  - o_{{\bf{n}}'} } \right) = \delta \left( {\theta _{\bf{n}}  - \theta '_{\bf{n}} } \right)\delta \left( {\varphi _{\bf{n}}  - \varphi '_{\bf{n}} } \right)/\sin \theta _{\bf{n}}$ acts as the solid-angle Dirac delta function. In the Schr\"{o}dinger picture, the Hamiltonian $\hat H = \int {d\omega } \;\hat H_\omega$ and the electric field operator ${\bf{\hat E}}\left( {\bf{r}} \right) = \int {d\omega } {\bf{\hat E}}_\omega  \left( {\bf{r}} \right) + {\rm H.c.}$ in the MLNF are such that:
\begin{align} \label{MLNF_HE}
 \hat H_\omega   &= \hbar \omega \left[ {\int {do_{\bf{n}} } {\bf{\hat g}}_{\omega s}^\dag  \left( {\bf{n}} \right) \cdot {\bf{\hat g}}_{\omega s} \left( {\bf{n}} \right) + \int_V {d^3 {\bf{r}}\,} \sum\nolimits_\nu  {{\bf{\hat f}}_{\omega \nu }^\dag  \left( {\bf{r}} \right) \cdot {\bf{\hat f}}_{\omega \nu } \left( {\bf{r}} \right)} } \right], \nonumber \\ 
 {\bf{\hat E}}_\omega  \left( {\bf{r}} \right) &= \int {do_{\bf{n}} } {\cal F}_{\omega s} \left( {\left. {\bf{r}} \right|{\bf{n}}} \right) \cdot {\bf{\hat g}}_{\omega s} \left( {\bf{n}} \right) + \int_V {d^3 {\bf{r}}'\,} \sum\nolimits_\nu  {{\cal G}_{\omega \nu } \left( {\left. {\bf{r}} \right|{\bf{r}}'} \right) \cdot {\bf{\hat f}}_{\omega \nu } \left( {{\bf{r}}'} \right)} 
\end{align}
where $do_{\bf{n}} = \sin \theta _{\bf{n}} d\theta _{\bf{n}} d\varphi _{\bf{n}}$ represents the differential solid angle. Scattering polaritons are integrated over the full solid angle to capture asymptotic free-space waves incident from spatial infinity, whereas the material polaritons ($\nu = e,m$) are integrated over the localized volume of the object to account for internal quantum noise sources. The spatial structure of the field expansion is governed by the scattering ($s$), electric ($e$), and magnetic ($m$)  dyadic kernels, explicitly given by:
\begin{equation} \label{Kernels}
{\cal F}_{\omega s} \left( {\left. {\bf{r}} \right|{\bf{n}}} \right) = {\cal F}_\omega  \left( {\left. {\bf{r}} \right|{\bf{n}}} \right)\sqrt {\frac{{\hbar k_\omega ^3 }}{{16\pi ^3 \varepsilon _0 }}} ,\quad \quad \begin{array}{*{20}l}
  {\cal G}_{\omega e} \left( {\left. {\bf{r}} \right|{\bf{r}}'} \right) &=& {\cal G}_\omega  \left( {\left. {\bf{r}} \right|{\bf{r}}'} \right) i\sqrt { \displaystyle \frac{{\hbar k_\omega ^4 }}{{\pi \varepsilon _0 }}{\mathop{\rm Im}\nolimits} \left[ {\varepsilon _\omega  \left( {{\bf{r}}'} \right)} \right]},  \\
  {\cal G}_{\omega m} \left( {\left. {\bf{r}} \right|{\bf{r}}'} \right) &=& {\cal G}_\omega  \left( {\left. {\bf{r}} \right|{\bf{r}}'} \right) \displaystyle \frac{{ \times \mathord{\buildrel{\lower3pt\hbox{$\scriptscriptstyle\leftarrow$}} 
\over \nabla } _{{\bf{r}}'} }}{{ik_\omega  }}\sqrt {\frac{{\hbar k_\omega ^4 }}{{\pi \varepsilon _0 }}{\mathop{\rm Im}\nolimits} \left[ {\frac{{ - 1}}{{\mu _\omega  \left( {{\bf{r}}'} \right)}}} \right]} ,
\end{array}
\end{equation}
where $k_\omega   = \omega /c$ denotes the vacuum wavenumber. These kernels are built entirely upon two fundamental tensors of classical macroscopic electrodynamics: the modal dyadic ${\cal F}_\omega  \left( {\left. {\bf{r}} \right|{\bf{n}}} \right)$ and the dyadic Green's function ${{\cal G}_\omega  \left( {\left. {\bf{r}} \right|{\bf{r}}'} \right)}$ which are  determined by the following Helmholtz boundary-value problems subject to asymptotic conditions:
\begin{align} \label{GS}
\left[ {\left( {\nabla _{\bf{r}}  \times \frac{1}{{\mu _\omega  \left( {\bf{r}} \right)}}\nabla _{\bf{r}}  \times } \right) - k_\omega ^2 \varepsilon _\omega  \left( {\bf{r}} \right)} \right]{\cal F}_\omega  \left( {\left. {\bf{r}} \right|{\bf{n}}} \right) &=  0,   & 
{\cal F}_\omega  \left( {\left. {r{\bf{m}}} \right|{\bf{n}}} \right) & \mathop  \approx \limits_{r \to + \infty } e^{i\left( {k_\omega  {\bf{n}}} \right) \cdot \left( {r{\bf{m}}} \right)} {\cal I}_{\bf{n}}  + \frac{{e^{ik_\omega  r} }}{r}{\cal S}_\omega  \left( {{\bf{m}}\left| {\bf{n}} \right.} \right),  \nonumber \\
\left[ {\left( {\nabla _{\bf{r}}  \times \frac{1}{{\mu _\omega  \left( {\bf{r}} \right)}}\nabla _{\bf{r}}  \times } \right) - k_\omega ^2 \varepsilon _\omega  \left( {\bf{r}} \right)} \right]{\cal G}_\omega  \left( {\left. {\bf{r}} \right|{\bf{r}}'} \right) & = \delta \left( {{\bf{r}} - {\bf{r}}'} \right){\cal I}, &
{\cal G}_\omega  \left( {\left. {r{\bf{m}}} \right|{\bf{r}}'} \right) & \mathop  \approx \limits_{r \to + \infty } \frac{{e^{ik_\omega  r} }}{r}{\cal W}_\omega  \left( {\left. {\bf{m}} \right|{\bf{r}}'} \right).
\end{align}
Here, $\bf m$ is a radial unit vector, and the symbol $\mathop \approx \limits_{r \to + \infty }$ isolates the leading-order term in the far-field expansion. Additionally, ${\cal S}_\omega \left( {{\bf{m}}\left| {\bf{n}} \right.} \right)$ represents the standard scattering dyadic \cite{Krist1}, whereas ${\cal W}_\omega \left( {{\bf{m}}\left| {\bf{r}} \right.} \right)$ characterizes the asymptotic far-field amplitude of the dyadic Green's function. Note that the modal dyadic ${\cal F}_\omega \left( {\left. {\bf{r}} \right|{\bf{n}}} \right)$ satisfies the right-transversality relation ${\cal F}_\omega \left( {\left. {\bf{r}} \right|{\bf{n}}} \right) \cdot {\bf{n}} =0$ and that it completely describes classical electromagnetic scattering since ${\bf{F}}_{\omega} \left( {\bf{r}} \right) = {\cal F}_{\omega} \left( {\left. {\bf{r}} \right|{\bf{n}}} \right) \cdot {\bf{e}}_{{\bf{n}}\perp }$ (with ${\bf{e}}_{{\bf{n}} \perp } \cdot {\bf{n}} = 0$) is the total field resulting from the scattering of an incoming plane wave of frequency $\omega$, direction ${\bf n}$, and transverse polarization ${\bf{e}}_{{\bf{n}} \perp }$. The asymptotic amplitude of the dyadic Green's function and the modal dyadic are linked by the exact relation 
\begin{equation} \label{W_F_relation}
{\cal W}_\omega  \left( {{\bf{m}}\left| {{\bf{r}}'} \right.} \right) = \frac{1}{{4\pi }}{\cal F}_\omega ^T \left( {\left. {{\bf{r}}'} \right| - {\bf{m}}} \right),
\end{equation}
which was originally derived in Ref.~\cite{Ciatt2024} in terms of scattering modes and it is central to the present analysis. Substituting this expression into the asymptotic behavior of the dyadic Green's function (fourth of Eqs.~(\ref{GS})) and invoking the reciprocity relation ${\cal G}_\omega ^T \left( {\left. {\bf{r}} \right|{\bf{r}}'} \right) = {\cal G}_\omega  \left( {\left. {{\bf{r}}'} \right|{\bf{r}}} \right)$, one readily obtains
\begin{equation} \label{G_F_relation}
{\cal G}_\omega  \left( {\left. {\bf{r}} \right|-r'{\bf{n}}} \right)\mathop  \approx \limits_{r' \to  + \infty } \frac{{e^{ik_\omega  r'} }}{{4\pi r'}}{\cal F}_\omega  \left( {\left. {\bf{r}} \right|  {\bf{n}}} \right),
\end{equation}
a relation that mathematically formalizes the well-known physical notion that scattering modes excited by plane waves propagating along the direction ${\bf n}$ can be identified with the fields generated by point dipole sources displaced to spatial infinity along the direction $-{\bf n}$. Finally, it is essential to note that the three dyadic kernels defining the spatial structure of the electric field operator are linked by the fundamental integral relation
\begin{equation} \label{Integral_Identity}
\int {do_{\bf{n}} } {\cal F}_{\omega s} \left( {\left. {\bf{r}} \right|{\bf{n}}} \right) \cdot {\cal F}_{\omega s}^{T*} \left( {\left. {{\bf{r}}'} \right|{\bf{n}}} \right) + \int_V {d^3 {\bf{R}}} \sum\limits_\nu  {{\cal G}_{\omega \nu } \left( {\left. {\bf{r}} \right|{\bf{R}}} \right) \cdot {\cal G}_{\omega \nu }^{T*} \left( {\left. {{\bf{r}}'} \right|{\bf{R}}} \right)}  = \frac{{\hbar k_\omega ^2 }}{{\pi \varepsilon _0 }}{\mathop{\rm Im}\nolimits} \left[ {{\cal G}_\omega  \left( {\left. {\bf{r}} \right|{\bf{r}}'} \right)} \right].
\end{equation}

\subsection{LNF}
The structural generality of the MLNF provides a natural mathematical pathway to recover the standard LNF as an exact reduction. This transition is rigorously established by considering the geometrical limit of a strictly unbounded dissipative medium, meaning that the object's volume extends to fill the entire space ($V \to \mathbb{R}^3$). Under this assumption, the modal dyadic ${\cal F}_\omega \left( {\left. {\bf{r}} \right|{\bf{n}}} \right)$ is governed by the homogeneous Helmholtz equation (the first of Eqs.~(\ref{GS})) across a globally lossy domain. Intuitively, sustaining a non-trivial wave in a medium that continuously absorbs energy everywhere would require an unbounded inward flux of power originating from spatial infinity. We formalize this physical intuition in Appendix~A, where, using the macroscopic energy balance, we rigorously demonstrate that any non-trivial solution is forced to diverge exponentially at large distances.  Since physically admissible scattering modes must remain bounded, the only globally regular solution is the trivial one: ${\cal F}_\omega \left( {\left. {\bf{r}} \right|{\bf{n}}} \right) \equiv 0$. Consequently, the scattering quantum kernel ${\cal F}_{\omega s} \left( {\left. {\bf{r}} \right|{\bf{n}}} \right)$ in Eqs.~(\ref{Kernels}) identically vanishes, and the scattering polaritons ${\bf{\hat g}}_{\omega s} \left( {\bf{n}} \right)$ systematically decouple from all local field observables. Because the entire asymptotic scattering sector becomes completely isolated and inaccessible within the Fock space, it can be formally omitted from the description of the interacting field-matter system without loss of generality. Hence, the spectral components of the Hamiltonian and of the electric field in Eqs.~(\ref{MLNF_HE}) naturally collapse to
\begin{equation} \label{LNF_H_E}
\hat H_\omega = \hbar \omega \int_{\mathbb{R}^3} d^3 {\bf{r}}\, \sum\nolimits_\nu {\bf{\hat f}}_{\omega \nu }^\dag \left( {\bf{r}} \right) \cdot {\bf{\hat f}}_{\omega \nu } \left( {\bf{r}} \right), \quad \quad \quad
{\bf{\hat E}}_\omega \left( {\bf{r}} \right) = \int_{\mathbb{R}^3} d^3 {\bf{r}}'\, \sum\nolimits_\nu {\cal G}_{\omega \nu } \left( \left. {\bf{r}} \right|{\bf{r}}' \right) \cdot {\bf{\hat f}}_{\omega \nu } \left( {{\bf{r}}'} \right),
\end{equation}
where the spatial integration over the internal material polaritons now spans the entirety of $\mathbb{R}^3$. Concurrently, the fundamental integral identity in Eq.~(\ref{Integral_Identity}) immediately reduces to
\begin{equation} \label{LNF_Integral_Identity}
\int_{\mathbb{R}^3} {d^3 {\bf{R}}} \sum\limits_\nu  {{\cal G}_{\omega \nu } \left( {\left. {\bf{r}} \right|{\bf{R}}} \right) \cdot {\cal G}_{\omega \nu }^{T*} \left( {\left. {{\bf{r}}'} \right|{\bf{R}}} \right)}  = \frac{{\hbar k_\omega ^2 }}{{\pi \varepsilon _0 }}{\mathop{\rm Im}\nolimits} \left[ {{\cal G}_\omega  \left( {\left. {\bf{r}} \right|{\bf{r}}'} \right)} \right].
\end{equation}
Equations~(\ref{LNF_H_E}) are exactly the foundational ones defining the macroscopic electric field and the Hamiltonian within the standard LNF. Furthermore, Eq.~(\ref{LNF_Integral_Identity}) corresponds exactly to the fundamental integral relation lacking any asymptotic surface term---routinely encountered throughout the LNF literature---which ensures the validity of the standard macroscopic field commutation relations. Thus, the LNF is naturally retrieved as the exact limiting case of the MLNF for unbounded lossy media.

\section{${\rm LNF}^{(\eta)}$ and limiting procedure}
Although the standard LNF is straightforwardly recovered as the unbounded-medium limit of the MLNF (as shown in Sec.~IIB), a profound structural asymmetry prevents the converse reduction, simply because the standard LNF intrinsically requires the absorption to be strictly non-vanishing at every point in space. Consequently, showing that the LNF framework can still recover the complete set of MLNF predictions for a finite object---including arbitrary quantum scattering processes---requires two fundamental ingredients: 1) the adoption of a tailored LNF framework where the vacuum surrounding the object is endowed with a fictitious absorption; and 2) a precise mathematical definition of convergence for this framework as the fictitious absorption vanishes. 

The first ingredient is implemented by considering the ${\rm LNF}^{(\eta)}$ framework governed by the spatial profiles
\begin{align} \label{eps_mu_eta}
\varepsilon _\omega ^{\left( \eta  \right)} \left( {\bf{r}} \right) =& \left\{ {\begin{array}{*{20}c}
   {\varepsilon _\omega ^V \left( {\bf{r}} \right),} & {{\bf{r}} \in V}  \\
   {1 + i\eta ,} & {{\bf{r}} \notin V}  \\
\end{array}} \right. &
\mu _\omega ^{\left( \eta  \right)} \left( {\bf{r}} \right) =& \left\{ {\begin{array}{*{20}c}
   {\mu _\omega ^V \left( {\bf{r}} \right),} & {{\bf{r}} \in V}  \\
   {1 + i\eta ,} & {{\bf{r}} \notin V,}  \\
\end{array}} \right.
\end{align}
which practically coincide with the exact physical parameters of Eqs.~(\ref{eps_mu}) apart from the fictitious uniform absorption outside the object's volume. Hereafter, the superscript $(\eta)$ appended to any MLNF quantity will denote its corresponding counterpart within the ${\rm LNF}^{(\eta)}$ framework, wherein the material polariton operators satisfy the macroscopic bosonic commutation relations
\begin{equation} \label{Comm_f^(eta)}
\left[ {{\bf{\hat f}}_{\omega \nu }^{\left( \eta  \right)} \left( {\bf{r}} \right),{\bf{\hat f}}_{\omega '\nu '}^{\left( \eta  \right)\dag } \left( {{\bf{r}}'} \right)} \right] = \delta \left( {\omega  - \omega '} \right)\delta _{\nu \nu '} \delta \left( {{\bf{r}} - {\bf{r}}'} \right){\cal I},
\end{equation}
valid for any arbitrary pair of points ${\bf{r}}$ and ${\bf{r}}'$ across the entire space $\mathbb{R}^3$.

Regarding the second ingredient, the precise mathematical sense in which the ${\rm LNF}^{(\eta)}$ framework reproduces the MLNF relies on preliminarily establishing a formal mapping between the two spaces. Specifically, one must exhibit a rigorous mapping, denoted hereafter by the symbol $\mapsto$, that associates any arbitrary MLNF state $\left| \Psi \right\rangle$ and operator $\hat{O}$ with their respective counterparts $\left| {\Psi^{\left( \eta \right)}} \right\rangle$ and $\hat{O}^{\left( \eta \right)}$ within the ${\rm LNF}^{(\eta)}$ framework, i.e., $\left| \Psi \right\rangle \mapsto \left| {\Psi^{\left( \eta \right)}} \right\rangle$ and $\hat{O} \mapsto \hat{O}^{\left( \eta \right)}$. However, the convergence of the theory cannot be established by taking the $\eta \to 0^+$ limit of these quantities, because, as will be explicitly demonstrated below for the scattering sector, certain mapped operators $\hat{O}^{(\eta)}$ converge  weakly to zero as $\eta \to 0^+$, failing to converge both strongly and in operator norm. Consequently, the convergence of the ${\rm LNF}^{(\eta)}$ to the MLNF must be assured by the convergence of arbitrary matrix elements, i.e., if the relation
\begin{equation} \label{Limiting_Procedure}
\lim_{\eta \to 0^+} \left\langle {\Phi^{\left( \eta \right)}} \right| \hat{O}^{\left( \eta \right)} \left| {\Psi^{\left( \eta \right)}} \right\rangle = \left\langle \Phi \right| \hat{O} \left| \Psi \right\rangle.
\end{equation}
holds for any pair of states $|\Phi\rangle$ and $|\Psi\rangle$, and for any operator $\hat{O}$. Note that this convergence criterion fundamentally departs from the standard weak operator topology established in functional analysis. Unlike the conventional weak convergence of $\hat{O}^{(\eta)}$ as $\eta \to 0^+$, which is defined through matrix elements taken between fixed states, here the states $| \Phi^{(\eta)} \rangle$ and $| \Psi^{(\eta)} \rangle$ are themselves generally dependent on the parameter $\eta$. From a physical perspective, this entails a systematic operational procedure. Given an arbitrary problem in macroscopic quantum electrodynamics, one first formalizes it within the MLNF and, leveraging the established mapping $\mapsto$, constructs the corresponding representation of all relevant states and observables within the ${\rm LNF}^{(\eta)}$ framework. If every physical prediction derived from this ${\rm LNF}^{(\eta)}$ formulation rigorously converges to the corresponding MLNF prediction in the limit $\eta \to 0^+$, the formal convergence of the two theoretical frameworks is fully demonstrated.

\section{Formal Mapping of the MLNF Scattering Sector into the ${\rm LNF}^{(\eta)}$ Framework}
As established in Sec.~III, demonstrating the exact convergence of the limiting procedure requires mapping arbitrary MLNF states into the ${\rm LNF}^{(\eta)}$ framework. Since the ${\rm LNF}^{(\eta)}$ Fock space intrinsically lacks a scattering sector, finding this mapping preliminarily necessitates identifying a specific subspace capable of emulating the exact scattering dynamics under the limiting procedure. Because the MLNF scattering sector is natively generated by polariton creation operators ${\bf{\hat g}}_{\omega s}^\dagger$ acting on the vacuum, we must identify a corresponding (albeit non-unique, see below) set of ${\rm LNF}^{(\eta)}$ operators ${\bf{\hat g}}_{\omega s}^{(\eta)}$ onto which the  MLNF scattering polaritons ${\bf{\hat g}}_{\omega s}$ are mapped, i.e. ${\bf{g}}_{\omega s} \left( {\bf{n}} \right) \mapsto {\bf{g}}_{\omega s}^{\left( \eta  \right)} \left( {\bf{n}} \right)$, in the mathematical sense of Eq.~(\ref{Limiting_Procedure}). It is worth noting that, by specializing Eq.~(\ref{Limiting_Procedure}) to the vacuum states $\left| \Phi  \right\rangle = \left| \Psi  \right\rangle = \left| 0  \right\rangle$ and the commutator $\hat O = [ {{\bf{\hat g}}_{\omega s} \left( {\bf{n}} \right),{\bf{\hat g}}_{\omega 's}^\dag \left( {{\bf{n}}'} \right)} ]$, we find that these ${\rm LNF}^{(\eta)}$ operators must satisfy the condition
\begin{equation} \label{Weak_Bosonic_Algebra}
\mathop {\lim }\limits_{\eta  \to 0^ +  } \left\langle {0^{(\eta)}} \right| \left[ {{\bf{\hat g}}_{\omega s}^{\left( \eta  \right)} \left( {\bf{n}} \right),{\bf{\hat g}}_{\omega 's}^{\left( \eta  \right)\dag } \left( {{\bf{n}}'} \right)} \right] \left| {0^{(\eta)}} \right\rangle = \delta \left( {\omega  - \omega '} \right)\delta \left( {o_{\bf{n}}  - o_{{\bf{n}}'} } \right){\cal I}_{\bf{n}},
\end{equation}
which precisely establishes their asymptotically bosonic nature.

To identify the operators ${\bf{\hat g}}_{\omega s}^{(\eta)}$, it is convenient to start from the expression of the ${\rm LNF}^{(\eta)}$ electric field  (see the second of Eqs.~(\ref{LNF_H_E})) and to split the spatial integration of the polariton operators into the regions inside and outside the object's volume, i.e.,
\begin{equation} \label{E_eta_split}
{\bf{\hat E}}_\omega ^{\left( \eta  \right)} \left( {\bf{r}} \right) = {\bf{\hat E}}_{\omega V}^{\left( \eta  \right)} \left( {\bf{r}} \right)+ {\bf{\hat E}}_{\omega s}^{\left( \eta  \right)} \left( {\bf{r}} \right),
\end{equation}
where
\begin{align}
{\bf{\hat E}}_{\omega V}^{\left( \eta  \right)} \left( {\bf{r}} \right) &= \int_V {d^3 {\bf{r}}'\,} \sum\nolimits_\nu  {{\cal G}_{\omega \nu }^{\left( \eta  \right)} \left( {\left. {\bf{r}} \right|{\bf{r}}'} \right) \cdot {\bf{\hat f}}_{\omega \nu }^{\left( \eta  \right)} \left( {{\bf{r}}'} \right)},  \nonumber \\ 
{\bf{\hat E}}_{\omega s}^{\left( \eta  \right)} \left( {\bf{r}} \right) &= \int_{\mathbb{R}^3 \setminus V} {d^3 {\bf{r}}'\,} \sum\nolimits_\nu  {{\cal G}_{\omega \nu }^{\left( \eta  \right)} \left( {\left. {\bf{r}} \right|{\bf{r}}'} \right) \cdot {\bf{\hat f}}_{\omega \nu }^{\left( \eta  \right)} \left( {{\bf{r}}'} \right)}. 
\end{align}

The term ${\bf{\hat E}}_{\omega V}^{\left( \eta \right)} \left( {\bf{r}} \right)$ represents the field generated by sources confined within the object's volume $V$, and its handling in the $\eta \to 0^+$ limit is straightforward. Indeed, the polariton operators ${\bf{\hat f}}_{\omega \nu }^{\left( \eta \right)}$ carry no intrinsic $\eta$-dependence, as this parameter appears explicitly only within the material permittivity and permeability defined in Eqs.~(\ref{eps_mu_eta}). Furthermore, for source points strictly inside the material (${\bf{r}}' \in V$), the ${\rm LNF}^{(\eta)}$ dyadic Green's function trivially converges to its MLNF counterpart, i.e., $\lim_{\eta \to 0^+} {\cal G}_{\omega \nu }^{\left( \eta \right)} \left( {\left. {\bf{r}} \right|{\bf{r}}'} \right) = {\cal G}_{\omega \nu } \left( {\left. {\bf{r}} \right|{\bf{r}}'} \right)$. Consequently, in the $\eta \to 0^+$ limit under consideration, we can set
\begin{equation}
{\bf{\hat E}}_{\omega V}^{\left( \eta \right)} \left( {\bf{r}} \right) = \int_V {d^3 {\bf{r}}'\,} \sum\nolimits_\nu  {{\cal G}_{\omega \nu } \left( {\left. {\bf{r}} \right|{\bf{r}}'} \right) \cdot {\bf{\hat f}}_{\omega \nu }^{\left( \eta \right)} \left( {{\bf{r}}'} \right)},
\end{equation}
which is formally identical to the material polariton contribution of the MLNF electric field [see the second of Eqs.~(\ref{MLNF_HE})], in the sense that the former is recovered from the latter strictly via the replacement ${\bf{\hat f}}_{\omega \nu } \left( {\bf{r}} \right) \mapsto {\bf{\hat f}}_{\omega \nu }^{\left( \eta \right)} \left( {\bf{r}} \right)$ for ${\bf{r}} \in V$.

The term ${\bf{\hat E}}_{\omega s}^{\left( \eta \right)} \left( {\bf{r}} \right)$ is entirely generated by the fictitious sources located in the unbounded region $\mathbb{R}^3 \setminus V$ surrounding the object, and it must represent the map into the ${\rm LNF}^{(\eta)}$ framework of the exact MLNF scattering contribution,
\begin{equation} \label{MLNF_sca}
{\bf{\hat E}}_{\omega s}  \left( {\bf{r}} \right) = \int do_{\bf{n}} \, {\cal F}_{\omega s} \left( \left. {\bf{r}} \right|{\bf{n}} \right) \cdot {\bf{\hat g}}_{\omega s} \left( {\bf{n}} \right).
\end{equation}
For this reason, ${\bf{\hat E}}_{\omega s}^{\left( \eta \right)} \left( {\bf{r}} \right)$ is hereafter referred to as the \emph{fictitious scattering field}. By substituting the explicit expressions for the kernels ${\cal G}_{\omega \nu }^{\left( \eta  \right)}\left( {\left. {\bf{r}} \right|{\bf{r}}'} \right)$ for source points ${\bf{r}}'$ within the fictitious vacuum [see Eqs.~(\ref{Kernels}) and (\ref{eps_mu_eta})], this field takes the form
\begin{equation} \label{Fic_sca_fie}
{\bf{\hat E}}_{\omega s}^{\left( \eta  \right)} \left( {\bf{r}} \right) = i\sqrt {\frac{{\hbar k_\omega ^4 }}{{\pi \varepsilon _0 }}\eta } \int_{\mathbb{R}^3 \setminus V} d^3 {\bf{r}}'\, {\cal G}_\omega ^{\left( \eta  \right)} \left( {\left. {\bf{r}} \right|{\bf{r}}'} \right) \cdot \left[ {{\bf{\hat f}}_{\omega e}^{\left( \eta  \right)} \left( {{\bf{r}}'} \right) - \frac{{\left( {{\cal I} \times \mathord{\buildrel{\lower3pt\hbox{$\scriptscriptstyle\leftarrow$}} \over \nabla } _{{\bf{r}}'} } \right)}}{{k_\omega  \sqrt {1 + \eta ^2 } }} \cdot {\bf{\hat f}}_{\omega m}^{\left( \eta  \right)} \left( {{\bf{r}}'} \right)} \right].
\end{equation}
In order to establish the mapping ${\bf{\hat E}}_{\omega s} \left( {\bf{r}} \right) \mapsto {\bf{\hat E}}_{\omega s}^{\left( \eta \right)} \left( {\bf{r}} \right)$, it is necessary to preliminarily demonstrate that, in the limit $\eta \to 0^+$, the fictitious scattering field is entirely generated by the polariton operators ${\bf{\hat f}}_{\omega \nu }^{\left( \eta  \right)} \left( {\bf{r}}' \right)$ located at spatial infinity ($r' \to \infty$). Indeed, by virtue of the asymptotic relation in Eq.~(\ref{G_F_relation}), this is the sole condition under which the dyadic Green's function entering Eq.~(\ref{Fic_sca_fie}) can yield the modal dyadic. This preliminary analysis of the sources warrants distinct mathematical scrutiny owing to two essential facts: (i) the operator ${\bf{\hat E}}_{\omega s}^{\left( \eta \right)} \left( {\bf{r}} \right)$, due to the prefactor $\sqrt{\eta}$, converges weakly to zero, failing to converge both strongly and in operator norm; and (ii) the asymptotic limit governing the confinement of the sources to spatial infinity must be parametrically tied to the $\eta \to 0^+$ limit, to avoid ambiguities arising from the possible non-commutativity of the two limits. A rigorous mathematical treatment addressing these subtleties is provided in Appendix~C where, by suitably partitioning $\mathbb{R}^3 \setminus V$ into spatial domains and analyzing the contributions of the sources contained within each domain to the quantum vacuum fluctuations of ${\bf{\hat E}}_{\omega s}^{\left( \eta \right)}$, it is strictly proven that: 1) the contribution arising from the sources contained within any finite subregion of $\mathbb{R}^3 \setminus V$ vanishes exactly as $\eta \to 0^+$; and 2) the effective sources are exclusively those located outside the expanding sphere of radius $(\eta^\alpha k_\omega)^{-1}$, with $0 < \alpha < 1$. Physically, the upper bound $\alpha < 1$ guarantees that the sphere expands strictly slower than the attenuation length $(\eta k_\omega)^{-1}$ induced by the fictitious absorption, thereby preventing the field from trivially vanishing. Indeed, since the finite-volume contribution already vanishes, setting $\alpha > 1$ would cause the exponential damping to completely suppress the sources at infinity. For definiteness, we set $\alpha = 1/2$ for the remainder of this work. Confining the integration in Eq.~(\ref{Fic_sca_fie}) strictly to this purely asymptotic region allows us to replace the ${\rm LNF}^{(\eta)}$ dyadic Green's function with its far-field asymptotic behavior, 
\begin{equation}
{\cal G}_\omega^{(\eta)}(\mathbf{r}|-r'\mathbf{n}) \mathop{\approx}\limits_{r' \to +\infty} \frac{e^{ik_\omega r'} e^{-\eta k_\omega r'}}{4\pi r'}  {\cal F}_\omega(\mathbf{r}|\mathbf{n}), 
\end{equation} 
a relation derived in Appendix~B, to be compared with its MLNF counterpart in Eq.~(\ref{G_F_relation}). As established above, this reduction is structurally pivotal, as it explicitly embeds the exact MLNF modal dyadic ${\cal F}_\omega$ into the ${\rm LNF}^{(\eta)}$ framework. Consequently, as detailed in Appendix~C, in the $\eta \to 0^+$ limit under consideration, the fictitious scattering field of Eq.~(\ref{Fic_sca_fie}) can effectively be expressed as
\begin{equation} \label{fic_sca_fie2}
{\bf{\hat E}}_{\omega s}^{\left( \eta  \right)} \left( {\bf{r}} \right) = \int {do_{\bf{n}} } {\cal F}_{\omega s} \left( {\left. {\bf{r}} \right|{\bf{n}}} \right) \cdot {\bf{g}}_{\omega s}^{\left( \eta  \right)} \left( {\bf{n}} \right),
\end{equation}
where we have introduced the operators
\begin{equation} \label{g^(eta)}
{\bf{g}}_{\omega s}^{\left( \eta  \right)} \left( {\bf{n}} \right) = \sqrt {\eta k_\omega  } {\cal I}_{\bf{n}}  \cdot \int_{\left( {\sqrt{\eta}  k_\omega  } \right)^{ - 1} }^{ + \infty } {dR \, R e^{ik_\omega  R} } e^{ - \eta k_\omega  R} \left[ {i{\bf{\hat f}}_{\omega e}^{\left( \eta  \right)} \left( { - R{\bf{n}}} \right) - {\bf{n}} \times {\bf{\hat f}}_{\omega m}^{\left( \eta  \right)} \left( { - R{\bf{n}}} \right)} \right].
\end{equation}
Note that the confinement of the fictitious scattering field sources to spatial infinity is mathematically encoded by restricting the integration defining the operators ${\bf{g}}_{\omega s}^{\left( \eta \right)} \left( {\bf{n}} \right)$ to the domain $R > (\sqrt{\eta} k_\omega)^{-1}$. 

At this stage, it is crucial to highlight that the fictitious scattering field in Eq.~(\ref{fic_sca_fie2}) exhibits exactly the same structure as the MLNF scattering contribution in Eq.~(\ref{MLNF_sca}). Such structural equivalence strongly suggests that the operators ${\bf{\hat g}}_{\omega s}^{\left( \eta \right)}$ are the sought-after ${\rm LNF}^{(\eta)}$ operators onto which the true MLNF polariton scattering operators ${\bf{\hat g}}_{\omega s}$ are mapped, in the sense of the limiting procedure introduced in Sec.~III. Although the rigorous justification of this fact will emerge as a consequence of the fundamental result discussed in the next section, it is crucial to note here that the mapping ${\bf{g}}_{\omega s} \left( {\bf{n}} \right) \mapsto {\bf{g}}_{\omega s}^{\left( \eta  \right)} \left( {\bf{n}} \right)$ agrees with the necessary condition given in Eq.~(\ref{Weak_Bosonic_Algebra}). Indeed, as shown in Appendix~C, the commutator evaluates to
\begin{equation} \label{Comm_g^(eta)}
\left[ {{\bf{\hat g}}_{\omega s}^{\left( \eta  \right)} \left( {\bf{n}} \right),{\bf{\hat g}}_{\omega 's}^{\left( \eta  \right)\dag } \left( {{\bf{n}}'} \right)} \right] = \Lambda \left( \eta  \right)\delta \left( {\omega  - \omega '} \right)\delta \left( {o_{\bf{n}}  - o_{{\bf{n}}'} } \right){\cal I}_{\bf{n}} ,
\end{equation}
where $\Lambda \left( \eta  \right) = e^{ - 2\sqrt \eta  }$, demonstrating that the relation 
\begin{equation} \label{Lambda}
\lim_{\eta \to 0^+} \Lambda(\eta) = 1
\end{equation} 
exactly guarantees the proper asymptotic bosonic nature of the operators ${\bf{\hat g}}_{\omega s}^{\left( \eta \right)}$. Furthermore, since the operators ${\bf{\hat g}}_{\omega s}^{\left( \eta \right)}$ in Eq.~(\ref{g^(eta)}) are constructed exclusively from material polaritonic operators ${\bf{\hat f}}_{\omega \nu}^{\left( \eta \right)}$ associated with the spatial region strictly outside the volume $V$, they trivially commute with the creation operators ${\bf{\hat f}}_{\omega' \nu}^{\left( \eta \right)\dag}$ pertaining to the volume $V$, yielding
\begin{equation} \label{Comm_g,f^(eta)}
\left[ {{\bf{\hat g}}_{\omega s}^{\left( \eta \right)} \left( {\bf{n}} \right), {\bf{\hat f}}_{\omega' \nu}^{\left( \eta \right)\dag} \left( {\bf{r}} \right)} \right] = 0, \quad {\bf{r}} \in V.
\end{equation}

\section{Asymptotic convergence of the ${\rm LNF}^{(\eta)}$ to the MLNF}

As discussed in Sec.~III, demonstrating the asymptotic convergence of the ${\rm LNF}^{(\eta)}$ to the MLNF preliminarily requires defining a mapping $\mapsto$ and subsequently proving the convergence of the corresponding matrix elements. In the present section, we systematically fulfill both of these requirements and analyze an explicit example to elucidate the mechanism underlying the convergence.

\subsection{The fundamental mapping $\mapsto$}
To construct the mapping $\mapsto$, we leverage the analysis of Sec.~III, which established that in the $\eta \to 0^+$ limit, the ${\rm LNF}^{(\eta)}$ electric field can be expressed as
\begin{equation}
{\bf{\hat E}}_\omega ^{\left( \eta  \right)} \left( {\bf{r}} \right) = \int {do_{\bf{n}} } {\cal F}_{\omega s} \left( {\left. {\bf{r}} \right|{\bf{n}}} \right) \cdot {\bf{\hat g}}_{\omega s}^{\left( \eta  \right)} \left( {\bf{n}} \right) + \int_V {d^3 {\bf{r}}'\,} \sum\nolimits_\nu  {{\cal G}_{\omega \nu } \left( {\left. {\bf{r}} \right|{\bf{r}}'} \right) \cdot {\bf{\hat f}}_{\omega \nu }^{\left( \eta  \right)} \left( {{\bf{r}}'} \right)}. 
\end{equation}
Crucially, this ${\rm LNF}^{(\eta)}$ operator is formally recovered from the MLNF electric field operator [second of Eqs.~(\ref{MLNF_HE})] strictly by replacing the underlying MLNF polaritonic operators with their ${\rm LNF}^{(\eta)}$ counterparts. Building upon this formal replacement rule, and observing that the Fock space and the entire operator algebra of the MLNF are generated by the action of polaritonic creation operators on the vacuum state, we define the mapping $\mapsto$ through the fundamental substitution procedure
\begin{equation} \label{Fund_map}
{\bf{\hat{g}}}_{\omega s} \left( {\bf{n}} \right) \mapsto {\bf{\hat{g}}}_{\omega s}^{(\eta)} \left( {\bf{n}} \right), \quad \quad \quad
{\bf{\hat{f}}}_{\omega \nu} \left( {\bf{r}} \right) \mapsto {\bf{\hat{f}}}_{\omega \nu}^{(\eta)} \left( {\bf{r}} \right), \quad \quad \quad 
\left| 0 \right\rangle \mapsto \left| 0^{(\eta)} \right\rangle.
\end{equation}
The exhaustive mathematical formalization of this structural correspondence and the associated notational framework are rigorously detailed in Appendix~D. To outline the underlying concept, let us consider an arbitrary state $\left| \Psi \right\rangle$ and an operator $\hat{O}$ within the MLNF framework, which can be expanded as

\begin{equation} \label{MLNF_sta_ope}
\begin{aligned}
\left| \Psi \right\rangle &= \sum\limits_{p,q,r = 0}^\infty \int d\Xi_{pqr} \, \Psi_{pqr} \left( {\Xi_{pqr}} \right) \left| {\Xi_{pqr}} \right\rangle, \\ 
\hat{O} &= \sum\limits_{p,q,r = 0}^\infty \sum\limits_{p',q',r' = 0}^\infty \int d\Xi_{pqr} \int d\Xi'_{p'q'r'} \, O\left( {\Xi_{pqr}, \Xi'_{p'q'r'}} \right) \left| {\Xi_{pqr}} \right\rangle \left\langle {\Xi'_{p'q'r'}} \right|.
\end{aligned}
\end{equation}
Here, the generic basis vectors 
\begin{equation}
\left| {\Xi _{pqr} } \right\rangle  = \left| {\xi _1^s , \ldots ,\xi _p^s ,\xi _1^e , \ldots ,\xi _q^e ,\xi _1^m , \ldots ,\xi _r^m } \right\rangle
\end{equation}
contain exactly $p$ scattering polaritons ($\xi ^s  = \left( {\omega ,{\bf{n}},\lambda } \right)$), $q$ electric polaritons and $r$ magnetic polaritons ($\xi ^\nu   = \left( {\omega ,{\bf{r}},\lambda } \right)$), which are generated from the MLNF vacuum $\left| 0 \right\rangle$ through the corresponding repeated application of the scalar creation operators ${\bf{e}}_{{\bf{n}}\lambda } \cdot {\bf{g}}_{\omega s}^\dag \left( {\bf{n}} \right)$ and ${\bf{u}}_\lambda \cdot {\bf{f}}_{\omega \nu }^\dagger \left( {\bf{r}} \right)$ (where ${\bf{e}}_{{\bf{n}}\lambda }$ and ${\bf{u}}_\lambda$ denote the polarization and spatial unit vectors, respectively). The MLNF state and operator of Eqs.~(\ref{MLNF_sta_ope}) are mapped by $\mapsto$ into the ${\rm LNF}^{(\eta)}$ state and operator
\begin{equation} \label{LNF^(eta)_sta_ope}
\begin{aligned}
\left| {\Psi^{\left( \eta \right)}} \right\rangle &= \sum\limits_{p,q,r = 0}^\infty \int d\Xi_{pqr} \, \Psi_{pqr} \left( {\Xi_{pqr}} \right) \left| {\Xi_{pqr}^{\left( \eta \right)}} \right\rangle, \\ 
\hat{O}^{\left( \eta \right)} &= \sum\limits_{p,q,r = 0}^\infty \sum\limits_{p',q',r' = 0}^\infty \int d\Xi_{pqr} \int d\Xi'_{p'q'r'} \, O\left( {\Xi_{pqr}, \Xi'_{p'q'r'}} \right) \left| {\Xi_{pqr}^{\left( \eta \right)}} \right\rangle \left\langle {\Xi'}_{p'q'r'}^{\left( \eta \right)} \right|,
\end{aligned}
\end{equation}
where the mapped vectors $| {\Xi_{pqr}^{\left( \eta \right)}} \rangle$ are constructed precisely through the replacement rules outlined in Eq.~(\ref{Fund_map}), namely via the repeated application of the scalar operators ${\bf{e}}_{{\bf{n}}\lambda } \cdot {\bf{\hat{g}}}_{\omega s}^{\left( \eta \right)\dag} \left( {\bf{n}} \right)$ and ${\bf{u}}_\lambda \cdot {\bf{\hat{f}}}_{\omega \nu }^{\left( \eta \right)\dag} \left( {\bf{r}} \right)$ onto the ${\rm LNF}^{(\eta)}$ vacuum state $\left| 0^{(\eta)} \right\rangle$. It is worth emphasizing that, to strictly preserve the physical content of the system, the mapping is explicitly constructed to act solely on the basis vectors $\left| {\Xi _{pqr} } \right\rangle$. Consequently, the scalar expansion coefficients ($\Psi_{pqr}$ and $O$)---which encode state preparation and intrinsic observables---remain rigorously unaltered, ensuring that the mapping translates only the theoretical framework without modifying the underlying physical scenario.

\subsection{Asymptotic convergence of the matrix elements}
We are now in a position to establish the central result of this work: the asymptotic convergence of arbitrary matrix elements, as stated in Eq.~(\ref{Limiting_Procedure}). While the rigorous mathematical proof is deferred to Appendix~D, the underlying mechanism driving this global convergence can be transparently outlined here. The crucial insight is that any mapped matrix element $\langle \Phi^{(\eta)} | \hat O^{(\eta)} | \Psi^{(\eta)} \rangle$ ultimately expands into a linear combination of terms of the form
\begin{equation} \label{2_inner}
\left\langle \tilde{\Xi}_{PQR}^{(\eta)} \middle| \Xi_{pqr}^{(\eta)} \right\rangle \left\langle \Xi^{\prime (\eta)}_{p'q'r'} \middle| \tilde{\Xi}^{\prime (\eta)}_{P'Q'R'} \right\rangle,
\end{equation}
which factorize into two inner products between the mapped basis vectors (naturally arising from sandwiching the operator on the left and right with the states). Crucially, the scalar coefficients weighting this expansion are strictly independent of $\eta$; by construction [see Eqs.~(\ref{LNF^(eta)_sta_ope})], they coincide exactly with the expansion coefficients of the corresponding MLNF matrix element $\langle \Phi | \hat O | \Psi \rangle$. The systematic evaluation of the inner products in Eq.~(\ref{2_inner}) relies entirely on the commutation relations governing the mapped polaritonic operators. As previously established, while the internal material operators ${\bf{\hat{f}}}_{\omega \nu}^{\left( \eta \right)}$ inherently maintain exact bosonic commutators [see Eq.~(\ref{Comm_f^(eta)})], the fictitious scattering operators ${\bf{\hat{g}}}_{\omega s}^{\left( \eta \right)}$ asymptotically recover the canonical bosonic commutation relations in the limit $\eta \to 0^+$ [see Eq.~(\ref{Comm_g^(eta)})]. Combining these properties with the exact mutual commutativity between  ${\bf{\hat{g}}}_{\omega s}^{(\eta)}$ and ${\bf{\hat{f}}}_{\omega \nu}^{(\eta)\dagger}$ [see Eq.~(\ref{Comm_g,f^(eta)})], the mapped scalar products asymptotically recover the exact orthonormality of the original MLNF basis space, yielding 
\begin{equation}
\lim_{\eta \to 0^+} \left\langle \Xi^{\prime (\eta)}_{p'q'r'} \middle| \Xi_{pqr}^{(\eta)} \right\rangle = \left\langle \Xi'_{p'q'r'} \middle| \Xi_{pqr} \right\rangle.
\end{equation}
Because the scalar expansion coefficients are rigorously preserved by the mapping, thus identically matching their MLNF counterparts, this asymptotic restoration of the basis orthonormality acts as the necessary and sufficient algebraic mechanism ensuring the exact convergence of the full matrix elements, thus yielding the fundamental result
\begin{equation}
\lim_{\eta \to 0^+} \left\langle {\Phi^{\left( \eta \right)}} \right| \hat{O}^{\left( \eta \right)} \left| {\Psi^{\left( \eta \right)}} \right\rangle = \left\langle \Phi \right| \hat{O} \left| \Psi \right\rangle.
\end{equation}

\subsection{An explicit example}
To elucidate the mechanism underlying this convergence, we now consider an explicit physical example. Let us first consider two specific states $|\Psi^i\rangle$ (with $i=1,2$) defined within the exact MLNF framework as
\begin{equation}
\left| {\Psi ^i } \right\rangle = \int d\omega \int do_{\bf{n}} \int d\Omega \int_V d^3 {\bf{R}} \, {\bf{\hat g}}_{\omega s}^\dag \left( {\bf{n}} \right) \cdot {\cal J}_{\omega \Omega }^i \left( {{\bf{n}},{\bf{R}}} \right) \cdot {\bf{\hat f}}_{\Omega e}^\dag \left( {\bf{R}} \right)\left| 0 \right\rangle.
\end{equation}
Physically, these vectors represent bipartite continuous-mode states comprising a single incident free-space photon ($s$-polariton) and a single localized material electric excitation ($e$-polariton). These two excitations are distributed, mathematically correlated, and possibly entangled through the left-transverse tensor ${\cal J}_{\omega \Omega }^i$ (i.e. ${\bf{n}} \cdot {\cal J}_{\omega \Omega }^i = 0$). Next, we introduce a transition operator $\hat{O}$, defined as
\begin{equation}
\hat O = \left[ {{\bf{e}}_{{\bf{n}}_2 }  \cdot {\bf{\hat g}}_{\omega _2 s}^\dag  \left( {{\bf{n}}_2 } \right)} \right]\left[ {{\bf{u}}_2  \cdot {\bf{\hat f}}_{\Omega _2 e}^\dag  \left( {{\bf{R}}_2 } \right)} \right]\left| 0 \right\rangle \left\langle 0 \right|\left[ {{\bf{u}}_1  \cdot {\bf{\hat f}}_{\Omega _1 e} \left( {{\bf{R}}_1 } \right)} \right]\left[ {{\bf{e}}_{{\bf{n}}_1 }  \cdot {\bf{\hat g}}_{\omega _1 s} \left( {{\bf{n}}_1 } \right)} \right].
\end{equation}
From a physical standpoint, this operator describes a specific transition between two sharply defined joint excitations. Specifically, it maps between states characterized by perfectly localized material dipoles (at positions ${\bf{R}}_{1,2}$ with orientations ${\bf{u}}_{1,2}$) and specific asymptotic incoming radiation modes (with distinct propagation directions ${\bf{n}}_{1,2}$ and polarization vectors ${\bf{e}}_{{\bf{n}}_{1,2}}$). Evaluating the corresponding matrix element systematically factors into the product of two overlap amplitudes. Physically, this represents the probability amplitude of finding the initially distributed states overlapping with the sharply defined modes selected by $\hat{O}$, which yields the product of the corresponding projected scalar tensors:
\begin{equation}
\left\langle {\Psi ^2 } \right|\hat{O}\left| {\Psi ^1 } \right\rangle = \left[ {{\bf{e}}_{{\bf{n}}_2 } \cdot {\cal J}_{\omega _2 \Omega _2 }^2 \left( {{\bf{n}}_2 ,{\bf{R}}_2 } \right) \cdot {\bf{u}}_2 } \right]^* \left[ {{\bf{e}}_{{\bf{n}}_1 } \cdot {\cal J}_{\omega _1 \Omega _1 }^1 \left( {{\bf{n}}_1 ,{\bf{R}}_1 } \right) \cdot {\bf{u}}_1 } \right].
\end{equation}
Having established the exact MLNF physical description, we now transition to the corresponding ${\rm LNF}^{(\eta)}$ framework. Applying the fundamental substitution procedure governed by Eq.~(\ref{Fund_map}), the mapped analogs of these states and operator are directly constructed as
\begin{align}
\left| {\Psi ^{i\left( \eta \right)} } \right\rangle &= \int d\omega \int do_{\bf{n}} \int d\Omega \int_V d^3 {\bf{R}} \, {\bf{\hat g}}_{\omega s}^{\left( \eta \right)\dag } \left( {\bf{n}} \right) \cdot {\cal J}_{\omega \Omega }^i \left( {{\bf{n}},{\bf{R}}} \right) \cdot {\bf{\hat f}}_{\Omega e}^{\left( \eta \right)\dag } \left( {\bf{R}} \right)\left| {0^{\left( \eta \right)} } \right\rangle, \\
\hat O^{\left( \eta  \right)}  &= \left[ {{\bf{e}}_{{\bf{n}}_2 }  \cdot {\bf{\hat g}}_{\omega _2 s}^{\left( \eta  \right)\dag } \left( {{\bf{n}}_2 } \right)} \right]\left[ {{\bf{u}}_2  \cdot {\bf{\hat f}}_{\Omega _2 e}^{\left( \eta  \right)\dag } \left( {{\bf{R}}_2 } \right)} \right]\left| {0^{\left( \eta  \right)} } \right\rangle \left\langle {0^{\left( \eta  \right)} } \right|\left[ {{\bf{u}}_1  \cdot {\bf{\hat f}}_{\Omega _1 e}^{\left( \eta  \right)} \left( {{\bf{R}}_1 } \right)} \right]\left[ {{\bf{e}}_{{\bf{n}}_1 }  \cdot {\bf{\hat g}} _{\omega _1 s}^{\left( \eta  \right)} \left( {{\bf{n}}_1 } \right) } \right]
\end{align}
When computing the matrix element $\langle {\Psi ^{2( \eta )} } |\hat{O}^{( \eta )} | {\Psi ^{1( \eta )} } \rangle$, the calculation relies on applying Wick's theorem to fully contract the mapped operators down to the vacuum state. Because the mapped material polariton operators rigorously preserve the exact canonical algebra [see Eq.~(\ref{Comm_f^(eta)})], their internal contractions yield standard orthonormal delta distributions, while their exact commutativity with the scattering sector [see Eq.~(\ref{Comm_g,f^(eta)})] ensures that all cross-contractions identically vanish. Conversely, every contraction between a mapped scattering operator and its adjoint is governed by the commutator in Eq.~(\ref{Comm_g^(eta)}), thereby extracting an explicit macroscopic factor $\Lambda(\eta)$. Since the evaluated matrix element describes a transition involving exactly one scattering excitation in both the initial and final states, Wick's theorem dictates exactly two complete scattering contractions---one evaluating the overlap with the bra, and one with the ket. Consequently, the calculation systematically accumulates an overall scaling factor of $\Lambda^2(\eta)$, yielding
\begin{equation}
\left\langle {\Psi ^{2\left( \eta  \right)} } \right|\hat O^{\left( \eta  \right)} \left| {\Psi ^{1\left( \eta  \right)} } \right\rangle  = \Lambda ^2 \left( \eta  \right)\left[ {{\bf{e}}_{{\bf{n}}_2 }  \cdot {\cal J}_{\omega _2 \Omega _2 }^2 \left( {{\bf{n}}_2 ,{\bf{R}}_2 } \right) \cdot {\bf{u}}_2 } \right]^* \left[ {{\bf{e}}_{{\bf{n}}_1 }  \cdot {\cal J}_{\omega _1 \Omega _1 }^1 \left( {{\bf{n}}_1 ,{\bf{R}}_1 } \right) \cdot {\bf{u}}_1 } \right].
\end{equation}
Ultimately, by taking the $\eta \to 0^+$ asymptotic limit, the macroscopic structural factor strictly evaluates to unity ($\lim_{\eta \to 0^+} \Lambda(\eta) = 1$), thereby perfectly recovering the exact MLNF theoretical outcome:
\begin{equation}
\lim_{\eta \to 0^+} \left\langle {\Psi ^{2\left( \eta \right)} } \right|\hat{O}^{\left( \eta \right)} \left| {\Psi ^{1\left( \eta \right)} } \right\rangle = \left\langle {\Psi ^2 } \right|\hat{O}\left| {\Psi ^1 } \right\rangle.
\end{equation}

\section{Discussion}
Before concluding, it is instructive to highlight a few salient features of our findings and to directly compare our approach with the most relevant precursor study in the literature.

First, the material sector mapping, ${\bf{\hat{f}}}_{\omega \nu} \left( {\bf{r}} \right) \mapsto {\bf{\hat{f}}}_{\omega \nu}^{(\eta)} \left( {\bf{r}} \right)$ for ${\bf{r}} \in V$, constitutes a straightforward, one-to-one translation of the internal excitations. Because localized material polaritons are natively defined in both frameworks and strictly preserve their macroscopic bosonic algebra, this correspondence is structurally trivial, yet essential for the formal completeness of the procedure. Most importantly, this exact structural correspondence rigorously justifies a highly relevant fact: the historical success of using the standard LNF to model finite objects in the absence of external incident radiation (e.g., thermal emission or Casimir physics). In such strictly internal-source regimes, the dynamics are exclusively governed by the material sector. Because the internal macroscopic operator algebra is mathematically identical in both frameworks, traditional LNF evaluations in these scenarios are formally indistinguishable from MLNF calculations where $s$-polaritons are not involved. Consequently, previous literature naturally obtained correct predictions, as those models were implicitly operating within the complete physical framework.

Second, the scattering sector mapping, ${\bf{\hat{g}}}_{\omega s} \mapsto {\bf{\hat{g}}}_{\omega s}^{(\eta)}$, represents the most non-trivial aspect of the presented mapping procedure. To the best of our knowledge, this mapping effectively resolves a previously unaddressed challenge in the literature: rigorously representing incoming free-space radiation within the standard LNF. Indeed, we have explicitly demonstrated how an initial MLNF state populated by incident photons (i.e., $s$-polaritons) is mathematically translated into an ${\rm LNF}^{(\eta)}$ state comprising specific excitations of the fictitious background infinitely far from the object. However, while formally feasible, employing the ${\rm LNF}^{(\eta)}$ to evaluate quantum scattering observables is mathematically cumbersome due to the necessity of rigorously computing the $\eta \to 0^+$ asymptotic limit. Consequently, for scenarios involving external incident radiation, it is distinctly more practical to directly adopt the MLNF, which inherently provides an exact, limit-free description of these quantum scattering processes~\cite{Ciatt2025_QuaSca,Ciatb2025_ClaTun,Ciattoni2026_FarFie,Ciattoni2026_QuaOpt}.

Finally, it is essential to contextualize our treatment alongside the work in Ref.~\cite{Drezet2017_EquBet}, wherein Drezet establishes the equivalence between the LNF and his Hamiltonian approach. The latter serves as the seminal precursor to the MLNF framework employed here, with the notable distinction that the MLNF natively incorporates the magnetic response of the scattering object as well. Specifically, Ref.~\cite{Drezet2017_EquBet} identifies an effective photon field generated by infinitely distant noise currents within the LNF. The effort to prove this equivalence relies on the possibility of---to use his exact words---``removing the independent photonic degrees of freedom though without breaking unitarity and time symmetry.'' Consequently, the analysis was inherently restricted to internal observables, thereby omitting scenarios involving external incident radiation. A crucial distinguishing feature of our framework with respect to the model in Ref.~\cite{Drezet2017_EquBet} therefore consists in demonstrating that the LNF rigorously converges to the MLNF across all physical configurations, explicitly encompassing standard quantum optical scenarios involving external incident radiation. Furthermore, the equivalence established in Ref.~\cite{Drezet2017_EquBet} relies on the premise that the effective photon field identified within the LNF---to use his exact words---``is formally equivalent to the scattered photon field'' (which corresponds to the incident radiation in his Hamiltonian approach). However, this formal identification can be further strengthened by a rigorous mathematical justification, which is strictly necessary for two fundamental reasons. First, the identification involves the interplay of two distinct limit procedures: the vanishing of the fictitious absorption and the infinite expansion of the external source volume. Notably, Ref.~\cite{Drezet2017_EquBet} addresses the mutual interplay of these limits exclusively within the specific calculation of the electric field commutator, lacking a generalized treatment. Second, the effective photon field fails to converge in the norm of the Hilbert space, a crucial mathematical aspect that falls outside the primary scope of the analysis in Ref.~\cite{Drezet2017_EquBet}. Our present derivation overcomes both of these critical mathematical hurdles. First, our analytic limiting procedure structurally couples these two limits by using the fictitious absorption parameter $\eta$ to dictate the radius $(\sqrt{\eta} k_\omega)^{-1}$ of an expanding sphere, beyond which the effective sources generating the fictitious scattering field are strictly localized. Second, our demonstration of convergence is mathematically rigorous, as it is evaluated directly on the matrix elements---a mandatory methodological choice, given that specific operators, such as the fictitious scattering field, converge weakly to zero, failing to converge strongly and in norm.

\section{Conclusions}
In summary, we have provided a mathematically rigorous demonstration that the standard Langevin Noise Formalism (LNF), when augmented with a fictitious vacuum absorption, asymptotically converges to the Modified Langevin Noise Formalism (MLNF). To achieve this, we established a systematic mapping procedure that uniquely associates any exact MLNF state and physical observable with a formal counterpart in the LNF framework. Crucially, rather than relying on the direct asymptotic limit of the field operators themselves---a procedure that is mathematically ill-defined since the fictitious scattering operator converges weakly to zero, failing to converge strongly and in norm---our proof is firmly grounded in the exact convergence of the arbitrary matrix elements. Through a rigorous analysis of the field generated by the fictitious vacuum, we demonstrated that its effective sources are strictly localized at spatial infinity. This mathematical deduction naturally dictates the coupling between the infinite expansion of the external source volume and the vanishing absorption limit $\eta \to 0^+$, providing the formal mechanism to rigorously prove that the mapped LNF matrix elements exactly recover the MLNF predictions. 

This mathematical result formally validates the LNF limiting procedure across all physical regimes. It not only ensures complete theoretical consistency with established historical results concerning internal fluctuation-induced phenomena---such as Casimir forces and thermal emission---but also definitively extends the theoretical validity of the LNF to encompass arbitrary external quantum scattering processes. 

Ultimately, while the two formalisms are thus fundamentally reconciled, the MLNF emerges as the significantly more practical and structurally natural tool for modern quantum optics. By natively incorporating external incident radiation as independent degrees of freedom, entirely bypassing the need for cumbersome asymptotic evaluations, the MLNF provides a highly convenient, exact, and limit-free platform for describing macroscopic quantum electrodynamics in arbitrary complex environments.

\appendix

\section{Vanishing of the modal dyadic in unbounded dissipative media}
In this Appendix, we provide a rigorous mathematical proof that if the dissipative medium strictly fills the entire space $\mathbb{R}^3$, the modal dyadic ${\cal F}_\omega \left( {\left. {\bf{r}} \right|{\bf{n}}} \right)$ identically vanishes everywhere. To simplify the notation without loss of generality, we project the modal dyadic onto an arbitrary transverse polarization unit vector ${\bf{e}}_{{\bf{n}} \bot }$ (where ${\bf{n}} \cdot {\bf{e}}_{{\bf{n}} \bot } = 0$), defining the classical scattering mode field ${\bf{F}}_\omega  \left( {\bf{r}} \right) = {\cal F}_\omega  \left( {\left. {\bf{r}} \right|{\bf{n}}} \right) \cdot {\bf{e}}_{{\bf{n}} \bot }$. According to the first of Eqs.~(\ref{GS}), this field is a solution to the homogeneous macroscopic wave equation:
\begin{equation} \label{HelmF}
\left[ \nabla _{\bf{r}}  \times \left( \frac{1}{{\mu _\omega  \left( {\bf{r}} \right)}}\nabla _{\bf{r}}  \times \right) - k_\omega ^2 \varepsilon _\omega  \left( {\bf{r}} \right) \right]{\bf{F}}_\omega  \left( {\bf{r}} \right) = 0.
\end{equation}
By taking the scalar product of the complex conjugate of this equation with the field ${\bf{F}}_\omega$, integrating over a spherical volume $V_R$ of radius $R$ bounded by the surface $\partial V_R$, applying the divergence theorem, and extracting the imaginary part, we obtain the integral identity:
\begin{equation} \label{AppA_Identity}
{\mathop{\rm Im}\nolimits} \oint_{\partial V_R } {dS\;{\bf{u}}_r \cdot } \left[ {{\bf{F}}_\omega   \times \left( {\frac{1}{{\mu _\omega ^* }}\nabla  \times {\bf{F}}_\omega ^* } \right)} \right] = \int_{V_R } {d^3 {\bf{r}}} \left[ {k_\omega ^2 {\mathop{\rm Im}\nolimits} \left( {\varepsilon _\omega  } \right)\left| {{\bf{F}}_\omega  } \right|^2  + {\mathop{\rm Im}\nolimits} \left( {\frac{{ - 1}}{{\mu _\omega  }}} \right)\left| {\nabla  \times {\bf{F}}_\omega  } \right|^2 } \right],
\end{equation}
where ${\bf{u}}_r$ is the outward radial unit vector. We recognize this integral identity as precisely the macroscopic Poynting theorem for monochromatic fields, which enforces the macroscopic energy balance within the present formalism. The assumption of a strictly unbounded lossy medium implies that the dissipative parameters are bounded from below by a strictly positive constant $\gamma > 0$ globally, while the inverse permeability is bounded from above by a constant $M$:
\begin{equation} \label{AppA_Bounds}
{\mathop{\rm Im}\nolimits} \left( {\varepsilon _\omega  } \right) \ge \gamma, \quad \quad {\mathop{\rm Im}\nolimits} \left( {\frac{{ - 1}}{{\mu _\omega  }}} \right) \ge \gamma, \quad \quad \sup \frac{1}{{\left| {\mu _\omega  } \right|}} = M < \infty.
\end{equation}
To analyze the spatial behavior of the field, we introduce a positive-definite energy-like function $W(R)$ representing the integrated field intensity within the sphere of radius $R$:
\begin{equation} \label{AppA_W}
W\left( R \right) = \int_{V_R } {d^3 {\bf{r}}} \left[ {k_\omega ^2 \left| {{\bf{F}}_\omega  } \right|^2  + \left| {\nabla  \times {\bf{F}}_\omega  } \right|^2 } \right],
\end{equation}
whose derivative with respect to the radius $R$ is exactly the integral of the energy density over the spherical shell:
\begin{equation} \label{AppA_dW}
\frac{{dW}}{{dR}} = \oint_{\partial V_R } {dS\;} \left( {k_\omega ^2 \left| {{\bf{F}}_\omega  } \right|^2  + \left| {\nabla  \times {\bf{F}}_\omega  } \right|^2 } \right).
\end{equation}
Applying the lower bounds from Eqs.~(\ref{AppA_Bounds}) to the right-hand side of the integral identity in Eq.~(\ref{AppA_Identity}), we establish a strict lower bound for the volume term in terms of $W(R)$:
\begin{equation} \label{AppA_VolBound}
\int_{V_R } {d^3 {\bf{r}}} \left[ {k_\omega ^2 {\mathop{\rm Im}\nolimits} \left( {\varepsilon _\omega  } \right)\left| {{\bf{F}}_\omega  } \right|^2  + {\mathop{\rm Im}\nolimits} \left( {\frac{{ - 1}}{{\mu _\omega  }}} \right)\left| {\nabla  \times {\bf{F}}_\omega  } \right|^2 } \right] \ge \gamma \int_{V_R } {d^3 {\bf{r}}} \left[ {k_\omega ^2 \left| {{\bf{F}}_\omega  } \right|^2  + \left| {\nabla  \times {\bf{F}}_\omega  } \right|^2 } \right] = \gamma W\left( R \right).
\end{equation}
Because the medium is absorbing, the RHS of Eq.~(\ref{AppA_Identity}) is trivially non-negative. Consequently, the imaginary part of the surface integral on the LHS is likewise non-negative, which implies it is bounded from above by the modulus of the complex integral itself. We can therefore establish the following chain of inequalities:
\begin{eqnarray} \label{AppA_SurBound}
 {\mathop{\rm Im}\nolimits} \oint_{\partial V_R } {dS\;{\bf{u}}_r \cdot } \left[ {{\bf{F}}_\omega   \times \left( {\frac{1}{{\mu _\omega ^* }}\nabla  \times {\bf{F}}_\omega ^* } \right)} \right] &\le& \left| { \oint_{\partial V_R } {dS\;{\bf{u}}_r \cdot } \left[ {{\bf{F}}_\omega   \times \left( {\frac{1}{{\mu _\omega ^* }}\nabla  \times {\bf{F}}_\omega ^* } \right)} \right]} \right| \nonumber \\ 
 &\le& \oint_{\partial V_R } {dS\;} \left| {{\bf{u}}_r \cdot \left[ {{\bf{F}}_\omega   \times \left( {\frac{1}{{\mu _\omega ^* }}\nabla  \times {\bf{F}}_\omega ^* } \right)} \right]} \right| \nonumber \\ 
 &\le& \frac{1}{{k_\omega  }}\oint_{\partial V_R } {dS\;} \frac{{\left| {k_\omega  {\bf{F}}_\omega  } \right|\left| {\nabla  \times {\bf{F}}_\omega  } \right|}}{{\left| {\mu _\omega  } \right|}} \nonumber \\ 
 &\le& \frac{M}{{2k_\omega  }}\oint_{\partial V_R } {dS\;} \left( {k_\omega ^2 \left| {{\bf{F}}_\omega  } \right|^2  + \left| {\nabla  \times {\bf{F}}_\omega  } \right|^2 } \right) = \frac{M}{{2k_\omega  }}\frac{{dW}}{{dR}}, 
\end{eqnarray}
where in the last inequality we have utilized the upper bound $M$ from Eqs.~(\ref{AppA_Bounds}) along with the arithmetic-geometric mean inequality ($ab \le \frac{1}{2}(a^2 + b^2)$ with $a = |k_\omega {\bf{F}}_\omega|$ and $b = |\nabla \times {\bf{F}}_\omega|$), and subsequently recognized the shell integral from Eq.~(\ref{AppA_dW}). Finally, by concatenating the upper bound of the surface integral from Eq.~(\ref{AppA_SurBound}), the integral identity of Eq.~(\ref{AppA_Identity}), and the lower bound of the volume integral from Eq.~(\ref{AppA_VolBound}), we obtain the reverse Grönwall-type inequality:
\begin{equation}
\frac{dW}{dR} \ge \frac{2\gamma k_\omega}{M} W.
\end{equation}
Multiplying by the integrating factor $e^{- \frac{2\gamma k_\omega}{M} R}$ and integrating from an arbitrary initial radius $R_0$ to $R > R_0$, we arrive at the exact exponential growth condition:
\begin{equation}
W\left( R \right) \ge W\left( {R_0 } \right)e^{\frac{{2\gamma }}{M}k_\omega  \left( {R - R_0 } \right)}.
\end{equation}
This result demonstrates that if the field is non-trivial, so that $W(R_0) > 0$ for some $R_0$, its integrated energy $W(R)$ diverges exponentially as $R \to +\infty$. Such macroscopic exponential growth necessarily implies that the local field ${\bf{F}}_\omega(\mathbf{r})$ itself diverges exponentially along some radial direction at spatial infinity, which is physically inadmissible. Consequently, the only physical solution to Eq.~(\ref{HelmF}) is the trivial one, ${\bf{F}}_\omega(\mathbf{r}) = 0$. Because the choice of the transverse polarization vector ${\bf{e}}_{{\bf{n}} \bot }$ was arbitrary, the entire modal dyadic identically vanishes everywhere, ${\cal F}_\omega \left( {\left. {\bf{r}} \right|{\bf{n}}} \right) \equiv 0$.

\section{Asymptotic behavior of ${\cal G}_\omega^{(\eta)}(\mathbf{r}|\mathbf{r}')$ for sources at large distances}
In this Appendix, we explicitly characterize the asymptotic behavior of the ${\rm LNF}^{(\eta)}$ dyadic Green's function ${\cal G}_\omega^{(\eta)}(\mathbf{r}|\mathbf{r}')$ for $r' \to +\infty$, which, having established that the effective sources are strictly confined to spatial infinity, governs the fictitious scattering field ${\bf{\hat E}}_{\omega s}^{\left( \eta  \right)} \left( {\bf{r}} \right)$ (see Eq.~(\ref{Fic_sca_fie})). Let ${\cal G}_{\omega0}(\mathbf{r}|\mathbf{r}')$ be the standard dyadic Green's function for pure vacuum, and ${\cal G}_{\omega0}^{(\eta)}(\mathbf{r}|\mathbf{r}')$ be the dyadic Green's function for a uniform fictitious dissipative background defined by $\varepsilon = \mu = 1+i\eta$. As is straightforward to demonstrate, the boundary-value problems governing the MLNF dyadic Green's function ${\cal G}_\omega(\mathbf{r}|\mathbf{r}')$ and the ${\rm LNF}^{(\eta)}$ dyadic Green's function ${\cal G}_\omega^{(\eta)}(\mathbf{r}|\mathbf{r}')$ are strictly equivalent to the following Lippmann-Schwinger integral equations:
\begin{equation} \label{AppB_LS}
\begin{aligned}
{\cal G}_\omega(\mathbf{r}|\mathbf{r}') &= {\cal G}_{\omega0}(\mathbf{r}|\mathbf{r}') - \int_V d^3\mathbf{R} \, {\cal G}_{\omega0}(\mathbf{r}|\mathbf{R}) \cdot \left[ K_\omega(\mathbf{R}) {\cal G}_\omega(\mathbf{R}|\mathbf{r}') \right], \\
{\cal G}_\omega^{(\eta)}(\mathbf{r}|\mathbf{r}') &= {\cal G}_{\omega0}^{(\eta)}(\mathbf{r}|\mathbf{r}') - \int_V d^3\mathbf{R} \, {\cal G}_{\omega0}^{(\eta)}(\mathbf{r}|\mathbf{R}) \cdot \left[ K_\omega^{(\eta)}(\mathbf{R}) {\cal G}_\omega^{(\eta)}(\mathbf{R}|\mathbf{r}') \right],
\end{aligned}
\end{equation}
where the respective macroscopic scattering operators are defined by the differential actions
\begin{equation} \label{AppB_K}
\begin{aligned}
K_\omega(\mathbf{R}) &= \nabla_{\mathbf{R}} \times \left[ \frac{1}{\mu_\omega(\mathbf{R})} - 1 \right] \nabla_{\mathbf{R}} \times - \, k_\omega^2 \left[\varepsilon_\omega(\mathbf{R}) - 1\right], \\
K_\omega^{(\eta)}(\mathbf{R}) &= \nabla_{\mathbf{R}} \times \left[ \frac{1}{\mu_\omega^{(\eta)}(\mathbf{R})} - \frac{1}{(1+i\eta)} \right] \nabla_{\mathbf{R}} \times - \, k_\omega^2 \left[ \varepsilon_\omega^{(\eta)}(\mathbf{R}) - (1+i\eta) \right].
\end{aligned}
\end{equation}
Note that the spatial integrations are strictly confined to the object's finite volume $V$, since the scattering operators $K_\omega$ and $K_\omega^{(\eta)}$ identically vanish outside this region. We now examine the far-field limit of Eqs.~(\ref{AppB_LS}) as the observation point moves to spatial infinity, setting $\mathbf{r} = r\mathbf{m}$ with $r \to +\infty$. In this regime, the unperturbed dyadic Green's functions exhibit the well-known asymptotic behaviors
\begin{equation} \label{AppB_G0}
\begin{aligned}
{\cal G}_{\omega0}(r\mathbf{m}|\mathbf{r}') &\mathop \approx \limits_{r \to +\infty} \frac{e^{ik_\omega r}}{4\pi r} e^{-ik_\omega\mathbf{m}\cdot\mathbf{r}'} {\cal I}_{\mathbf{m}}, \\
{\cal G}_{\omega0}^{(\eta)}(r\mathbf{m}|\mathbf{r}') &\mathop \approx \limits_{r \to +\infty} \frac{e^{ik_\omega(1+i\eta)r}}{4\pi r} e^{-ik_\omega(1+i\eta)\mathbf{m}\cdot\mathbf{r}'} {\cal I}_{\mathbf{m}}.
\end{aligned}
\end{equation}
By inserting these expressions into the corresponding integral equations, the full dyadic Green's functions assume the asymptotic forms
\begin{equation} \label{AppB_G}
\begin{aligned}
{\cal G}_\omega(r\mathbf{m}|\mathbf{r}') &\mathop \approx \limits_{r \to +\infty} \frac{e^{ik_\omega r}}{r} {\cal W}_\omega(\mathbf{m}|\mathbf{r}'), \\
{\cal G}_\omega^{(\eta)}(r\mathbf{m}|\mathbf{r}') &\mathop \approx \limits_{r \to +\infty} \frac{e^{ik_\omega(1+i\eta)r}}{r} {\cal W}_\omega^{(\eta)}(\mathbf{m}|\mathbf{r}'),
\end{aligned}
\end{equation}
where the respective asymptotic amplitudes are explicitly given by
\begin{equation} \label{AppB_W}
\begin{aligned}
{\cal W}_\omega(\mathbf{m}|\mathbf{r}') &= \frac{1}{4\pi} \left\{ e^{-ik_\omega\mathbf{m}\cdot\mathbf{r}'} {\cal I}_{\mathbf{m}} - \int_V d^3\mathbf{R} \, e^{-ik_\omega\mathbf{m}\cdot\mathbf{R}} {\cal I}_{\mathbf{m}} \cdot \left[ K_\omega(\mathbf{R}) {\cal G}_\omega(\mathbf{R}|\mathbf{r}') \right] \right\}, \\
{\cal W}_\omega^{(\eta)}(\mathbf{m}|\mathbf{r}') &= \frac{1+i\eta}{4\pi} \left\{ e^{-ik_\omega(1+i\eta)\mathbf{m}\cdot\mathbf{r}'} {\cal I}_{\mathbf{m}} - \int_V d^3\mathbf{R} \, e^{-ik_\omega(1+i\eta)\mathbf{m}\cdot\mathbf{R}} {\cal I}_{\mathbf{m}} \cdot \left[ K_\omega^{(\eta)}(\mathbf{R}) {\cal G}_\omega^{(\eta)}(\mathbf{R}|\mathbf{r}') \right] \right\}.
\end{aligned}
\end{equation}
Since the spatial integration in the second of Eqs.~(\ref{AppB_W}) is strictly bounded to the finite volume $V$, the integral is a well-behaved analytic function of $\eta$. This allows us to rigorously evaluate its asymptotic expansion for $\eta \to 0^+$. By extracting the leading-order contribution, the  term $K_\omega^{(\eta)}(\mathbf{R}) {\cal G}_\omega^{(\eta)}(\mathbf{R}|\mathbf{r}') $ reduces exactly to its unperturbed counterpart $\ K_\omega(\mathbf{R}) {\cal G}_\omega(\mathbf{R}|\mathbf{r}')$, and the exponential factors revert to their real-phase form $e^{-ik_\omega \mathbf{m}\cdot\mathbf{R}}$. Comparing this result directly with the exact MLNF amplitude in the first of Eqs.~(\ref{AppB_W}), we establish that the two far-field amplitudes coincide up to a linear error in $\eta$:
\begin{equation} \label{AppB_W_Expansion}
{\cal W}_\omega^{(\eta)}(\mathbf{m}|\mathbf{r}') = {\cal W}_\omega(\mathbf{m}|\mathbf{r}') + {\cal O}(\eta).
\end{equation}

Finally, we evaluate the asymptotic behavior of the ${\rm LNF}^{(\eta)}$ dyadic Green's function for sources at large distances. By invoking the reciprocity relation ${\cal G}_\omega^{(\eta)T}(\mathbf{r}|\mathbf{r}') = {\cal G}_\omega^{(\eta)}(\mathbf{r}'|\mathbf{r})$ and setting the source position to $\mathbf{r}' = -r'\mathbf{n}$ (which dictates the outgoing direction $\mathbf{m} = -\mathbf{n}$ for the transposed variable), we substitute the amplitude expansion of Eq.~(\ref{AppB_W_Expansion}). We then exploit the exact identity ${\cal W}_\omega(-\mathbf{n}|\mathbf{r}) = \frac{1}{4\pi}{\cal F}_\omega^T(\mathbf{r}|\mathbf{n})$ deduced from Eq.~(\ref{W_F_relation}). Reversing the transpose to act on the spatial argument, we obtain the definitive asymptotic approximation for the fictitious dyadic Green's function:
\begin{equation} \label{AppB_Final_Asymptotic}
 {\cal G}_\omega^{(\eta)}(\mathbf{r}|-r'\mathbf{n}) 
 \mathop \approx \limits_{r' \to +\infty} \frac{e^{ik_\omega r'} e^{-\eta k_\omega r'}}{4\pi r'} \left[ {\cal F}_\omega(\mathbf{r}|\mathbf{n}) + {\cal O}(\eta) \right].
\end{equation}
Comparing this relation with Eq.~(\ref{G_F_relation}) explicitly reveals that the asymptotic behaviors of the two dyadic Green's functions differ primarily by the exponential damping factor $e^{-\eta k_\omega r'}$. Equation~(\ref{AppB_Final_Asymptotic}) is central to the present analysis, since it formally connects the asymptotic behavior of the ${\rm LNF}^{(\eta)}$ dyadic Green's function for sources at large distances to the exact MLNF modal dyadic ${\cal F}_\omega(\mathbf{r}|\mathbf{n})$.

\section{Rigorous mathematical identification of the sources generating the fictitious scattering field in the asymptotic limit}
In this appendix, we consider the fictitious scattering field operator ${\bf{\hat E}}_{\omega s}^{\left( \eta  \right)} \left( {\bf{r}} \right)$ defined in Eq.~(\ref{Fic_sca_fie}), and we rigorously demonstrate that, in the limit $\eta \to 0^+$, it is generated exclusively by infinitely distant sources ($R \to \infty$), providing a precise mathematical meaning to this statement. This analysis cannot be conducted directly on the operator ${\bf{\hat E}}_{\omega s}^{\left( \eta  \right)} \left( {\bf{r}} \right)$ because, as can be straightforwardly demonstrated using standard techniques of functional analysis, it weakly converges to the null operator in the limit $\eta \to 0^+$, while failing to converge both strongly and in operator norm. Consequently, to rigorously identify which spatial regions of polaritonic sources effectively contribute to ${\bf{\hat E}}_{\omega s}^{\left( \eta  \right)} \left( {\bf{r}} \right)$, one must necessarily analyze the quantum vacuum fluctuations of the field, namely $\left\langle {0^{\left( \eta  \right)} } \right|{\bf{\hat E}}_{\omega s}^{\left( \eta  \right)} \left( {\bf{r}} \right){\bf{\hat E}}_{\omega' s}^{\left( \eta  \right)\dag } \left( {{\bf{r}}'} \right)\left| {0^{\left( \eta  \right)} } \right\rangle$.

\subsection{Spatial partition of the fictitious scattering field sources and domain contributions}
In the existing literature, the vanishing fictitious absorption limit ($\eta \to 0^+$) and the asymptotic boundary limit for the source volume ($R \to \infty$) are conventionally treated as two independent operations. However, from a physical perspective, $\eta$ should act as the single regularization parameter governing the asymptotic behavior of the macroscopic sources and the overall convergence of the theory. Furthermore, given the singular nature of the procedure, analytical ambiguities related to the non-commutativity of multiple limits may arise. For these reasons, we explicitly couple the two limits by partitioning the space $\mathbb{R}^3 \setminus V$, occupied by the sources of the fictitious scattering field, into three mutually disjoint domains:
\begin{equation}
D_1 = \left( \mathbb{R}^3 \setminus V \right) \bigcap \left\{ R < \bar{R}_\omega \right\}, \quad
D_2 = \left\{ \bar{R}_\omega < R < R_\omega(\eta) \right\}, \quad
D_3 = \left\{ R > R_\omega(\eta) \right\},
\end{equation}
where ${\bf{R}}$ denotes the source position vector (with $R = |{\bf{R}}|$), while the fixed macroscopic radius $\bar{R}_\omega$ and the expanding boundary $R_\omega(\eta)$ are chosen such that 
\begin{equation} \label{Radii}
\bar{R}_\omega > {\rm diam}(V), \quad k_\omega \bar{R}_\omega \gg 1, \quad R_\omega(\eta) > \bar{R}_\omega, \quad \lim_{\eta \to 0^+} R_\omega(\eta) = +\infty.
\end{equation}
Here, $D_1$ is the macroscopically large, yet finite, region surrounding the object volume $V$, confined within the sphere of radius $\bar{R}_\omega$. Conversely, both $D_2$ and $D_3$ are located in the far field $R>\bar{R}_\omega$: $D_2$ defines an intermediate macroscopic spherical shell whose outer boundary expands dynamically as $\eta \to 0^+$, and $D_3$ constitutes the purely asymptotic region, which is strictly confined to spatial infinity in the limit $\eta \to 0^+$. Based on the spatial partition $\mathbb{R}^3 \setminus V = \bigcup_{j=1}^3 D_j$, the fictitious scattering field of Eq.~(\ref{Fic_sca_fie}) is accordingly decomposed into three contributions:
\begin{equation}
{\bf{\hat E}}_{\omega s}^{\left( \eta  \right)} \left( {\bf{r}} \right) = \sum_{j = 1}^3 {{\bf{\hat E}}_{\omega sj}^{\left( \eta  \right)} \left( {\bf{r}} \right)},
\end{equation}
where the field generated by each domain $D_j$ is given by
\begin{equation} \label{Ej}
{\bf{\hat E}}_{\omega sj}^{\left( \eta  \right)} \left( {\bf{r}} \right) = i\sqrt {\frac{{\hbar k_\omega ^4 }}{{\pi \varepsilon _0 }}\eta } \int_{D_j } {d^3 {\bf{R}}\,} \left[ {{\cal G}_\omega ^{\left( \eta  \right)} \left( {\left. {\bf{r}} \right|{\bf{R}}} \right) \cdot {\bf{\hat f}}_{\omega e}^{\left( \eta  \right)} \left( {\bf{R}} \right) - \frac{{{\cal G}_\omega ^{\left( \eta  \right)} \left( {\left. {\bf{r}} \right|{\bf{R}}} \right) \times \mathord{\buildrel{\lower3pt\hbox{$\scriptscriptstyle\leftarrow$}} \over \nabla } _{\bf{R}} }}{{k_\omega  \sqrt {1 + \eta ^2 } }} \cdot {\bf{\hat f}}_{\omega m}^{\left( \eta  \right)} \left( {\bf{R}} \right)} \right].
\end{equation}
The advantage of introducing the fixed sphere is that the far-field condition $k_\omega \bar{R}_\omega \gg 1$ explicitly ensures that, for $R > \bar{R}_\omega$, the dyadic Green's function and its curl are rigorously described by their asymptotic behavior for very distant sources, as derived in Appendix B [Eq.~(\ref{AppB_Final_Asymptotic})], namely
\begin{align}
 {\cal G}_\omega ^{\left( \eta  \right)} \left( {\left. {\bf{r}} \right|R{\bf{n}}} \right) &\mathop \approx \limits_{R \to  + \infty } \frac{{e^{ik_\omega  \left( {1 + i\eta } \right)R} }}{{4\pi R}}{\cal F}_\omega  \left( {\left. {\bf{r}} \right| - {\bf{n}}} \right), \nonumber \\ 
 {\cal G}_\omega ^{\left( \eta  \right)} \left( {\left. {\bf{r}} \right|R{\bf{n}}} \right) \times \mathord{\buildrel{\lower3pt\hbox{$\scriptscriptstyle\leftarrow$}} \over \nabla } _{\bf{R}} &\mathop \approx \limits_{R \to  + \infty } ik_\omega  \left( {1 + i\eta } \right)\frac{{e^{ik_\omega  \left( {1 + i\eta } \right)R} }}{{4\pi R}}{\cal F}_\omega  \left( {\left. {\bf{r}} \right| - {\bf{n}}} \right) \times {\bf{n}}.
\end{align}
Accordingly, in the far-field domains $D_2$ and $D_3$, these expressions allow us to simplify the field in Eq.~(\ref{Ej}). By introducing the macroscopic scattering kernel ${\cal F}_{\omega s}$ via the first of Eqs.~(\ref{Kernels}), switching to spherical coordinates for the spatial integration, and redefining the dummy angular integration variable as ${\bf{n}} \to -{\bf{n}}$, the field for $j=2,3$ can be conveniently expressed as
\begin{equation} \label{Fic_fiel_23}
 {\bf{\hat E}}_{\omega sj}^{\left( \eta  \right)} \left( {\bf{r}} \right) = \int {do_{\bf{n}} } {\cal F}_{\omega s} \left( {\left. {\bf{r}} \right|{\bf{n}}} \right) \cdot {\bf{g}}_{\omega sj}^{\left( \eta  \right)} \left( {\bf{n}} \right),
\end{equation}
where we have defined the operators
\begin{equation} \label{Def_g^eta_j}
{\bf{g}}_{\omega sj}^{\left( \eta  \right)} \left( {\bf{n}} \right) = \sqrt {\eta k_\omega  } {\cal I}_{\bf{n}}  \cdot \int_{D_j } {dR \, R} \, e^{ik_\omega  \left( {1 + i\eta } \right)R} \left[ {i{\bf{\hat f}}_{\omega e}^{\left( \eta  \right)} \left( { - R{\bf{n}}} \right) - \sqrt {\frac{{1 + i\eta }}{{1 - i\eta }}} {\bf{n}} \times {\bf{\hat f}}_{\omega m}^{\left( \eta  \right)} \left( { - R{\bf{n}}} \right)} \right],
\end{equation}
and the dyadic ${\cal I}_{\bf{n}}$ has been introduced to account for the right-transversality of the scattering kernel ${\cal F}_{\omega s}$, i.e., ${\cal F}_{\omega s} \left( {\left. {\bf{r}} \right|{\bf{n}}} \right) \cdot {\bf{n}} = 0$. 

At this stage, it is crucial to determine the commutation relations $[ {{\bf{\hat g}}_{\omega sj}^{\left( \eta  \right)} \left( {\bf{n}} \right),{\bf{\hat g}}_{\omega 'sj}^{\left( \eta  \right)\dag } \left( {{\bf{n}}'} \right)} ]$. By substituting the definition of these operators and expanding the commutator linearly, only the auto-commutators survive, since the cross-commutators between the electric ($\nu=e$) and magnetic ($\nu=m$) polaritonic operators identically vanish. This yields
\begin{align}
\left[ {{\bf{\hat g}}_{\omega sj}^{\left( \eta  \right)} \left( {\bf{n}} \right),{\bf{\hat g}}_{\omega 'sj}^{\left( \eta  \right)\dag } \left( {{\bf{n}}'} \right)} \right] &= \eta \sqrt {k_\omega  k_{\omega '} } \int_{D_j } {dR \, R} \, e^{ik_\omega  R} e^{ - \eta k_\omega  R} \int_{D_j } {dR' \, R'} \, e^{ - ik_{\omega '} R'} e^{ - \eta k_{\omega '} R'} \nonumber \\
&\quad \times {\cal I}_{\bf{n}}  \cdot \left\{ {\left[ {{\bf{\hat f}}_{\omega e}^{\left( \eta  \right)} \left( { - R{\bf{n}}} \right),{\bf{\hat f}}_{\omega 'e}^{\left( \eta  \right)\dag } \left( { - R'{\bf{n}}'} \right)} \right] - {\bf{n}} \times \left[ {{\bf{\hat f}}_{\omega m}^{\left( \eta  \right)} \left( { - R{\bf{n}}} \right),{\bf{\hat f}}_{\omega 'm}^{\left( \eta  \right)\dag } \left( { - R'{\bf{n}}'} \right)} \right] \times {\bf{n}}'} \right\} \cdot {\cal I}_{{\bf{n}}'}.
\end{align}
To proceed, we evaluate the core of the integral by applying the commutation relations for the material polariton operators given in Eq.~(\ref{Comm_f^(eta)}). The resulting spatial Dirac delta function is expressed in spherical coordinates along the integration rays, yielding $\delta ( - R{\bf{n}} + R'{\bf{n}}' ) = R^{ - 2} \delta (R - R')\delta (o_{\bf{n}}  - o_{{\bf{n}}'} )$. Exploiting the frequency delta to set $\omega' = \omega$ and the angular delta to enforce ${\bf{n}}' = {\bf{n}}$ within the tensorial part, the expression reduces to
\begin{align}
\left[ {{\bf{\hat g}}_{\omega sj}^{\left( \eta  \right)} \left( {\bf{n}} \right),{\bf{\hat g}}_{\omega 'sj}^{\left( \eta  \right)\dag } \left( {{\bf{n}}'} \right)} \right] &= \eta k_\omega \int_{D_j } {dR \, R} \, e^{ik_\omega  R} e^{ - \eta k_\omega  R} \int_{D_j } {dR' \, R'} \, e^{ - ik_\omega  R'} e^{ - \eta k_\omega  R'} \nonumber \\
&\quad \times \delta \left( {\omega  - \omega '} \right)\frac{{\delta \left( {R - R'} \right)}}{{R^2 }}\delta \left( {o_{\bf{n}}  - o_{{\bf{n}}'} } \right){\cal I}_{\bf{n}}  \cdot \left[ {{\cal I} - {\bf{n}} \times {\cal I} \times {\bf{n}}} \right] \cdot {\cal I}_{\bf{n}}.
\end{align}
The integration over $R'$ is trivially performed due to the radial delta function, thereby exactly canceling the $R^2$ denominator. For the tensorial part, applying the dyadic identity ${\bf{n}} \times {\cal I} \times {\bf{n}} = - {\cal I}_{\bf{n}}$ simplifies the core to ${\cal I} + {\cal I}_{\bf{n}}$, which upon projection from both sides with ${\cal I}_{\bf{n}}$ yields the contraction $2{\cal I}_{\bf{n}}$. The commutator thus simplifies to
\begin{equation} \label{Comm_g^eta_j}
\left[ {{\bf{\hat g}}_{\omega sj}^{\left( \eta  \right)} \left( {\bf{n}} \right),{\bf{\hat g}}_{\omega 'sj}^{\left( \eta  \right)\dag } \left( {{\bf{n}}'} \right)} \right] = \Lambda_{\omega j} \left( \eta  \right) \delta \left( {\omega  - \omega '} \right)\delta \left( {o_{\bf{n}}  - o_{{\bf{n}}'} } \right){\cal I}_{\bf{n}},
\end{equation}
where we have defined the domain-dependent factor 
\begin{equation} \label{Lamb_j}
\Lambda_{\omega j} (\eta) = 2\eta k_\omega \int_{D_j} dR \, e^{-2\eta k_\omega R} = 
\begin{cases}
\displaystyle 2\eta k_\omega \int_{\bar{R}_\omega}^{R_\omega(\eta)} dR \, e^{-2\eta k_\omega R} = e^{-2\eta k_\omega \bar{R}_\omega} - e^{-2\eta k_\omega R_\omega(\eta)}, & j = 2, \\[10pt]
\displaystyle 2\eta k_\omega \int_{R_\omega(\eta)}^\infty dR \, e^{-2\eta k_\omega R} = e^{-2\eta k_\omega R_\omega(\eta)}, & j = 3.
\end{cases}
\end{equation}

\subsection{Source localization via vacuum field fluctuations}
We now proceed to evaluate the vacuum expectation value of the field correlation. In this calculation, the cross-domain terms vanish because the three spatial domains are mutually disjoint and the material polariton operators ${\bf{\hat f}}_{\omega \nu }^{\left( \eta  \right)}$ are spatially uncorrelated (see Eq.~(\ref{Comm_f^(eta)})), thereby reducing the correlation exactly to the sum of the commutators over the individual domains:
\begin{equation} \label{Vac_flu}
\left\langle {0^{\left( \eta  \right)} } \right|{\bf{\hat E}}_{\omega s}^{\left( \eta  \right)} \left( {\bf{r}} \right){\bf{\hat E}}_{\omega' s}^{\left( \eta  \right)\dag } \left( {{\bf{r}}'} \right)\left| {0^{\left( \eta  \right)} } \right\rangle  = \sum_{j = 1}^3 {\left\langle {0^{\left( \eta  \right)} } \right|\left[ {{\bf{\hat E}}_{\omega sj}^{\left( \eta  \right)} \left( {\bf{r}} \right),{\bf{\hat E}}_{\omega' sj}^{\left( \eta  \right)\dag } \left( {{\bf{r}}'} \right)} \right]\left| {0^{\left( \eta  \right)} } \right\rangle }.
\end{equation}
Using the canonical commutation relations of the polariton operators in Eq.~(\ref{Comm_f^(eta)}), the field commutator for the domain $D_1$ is structurally expressed as
\begin{equation}
\begin{split}
\left[ {{\bf{\hat E}}_{\omega s1}^{\left( \eta  \right)} \left( {\bf{r}} \right),{\bf{\hat E}}_{\omega' s1}^{\left( \eta  \right)\dag } \left( {{\bf{r}}'} \right)} \right] &= \delta \left( {\omega  - \omega '} \right)\eta \frac{{\hbar k_\omega ^4 }}{{\pi \varepsilon _0 }}\int_{D_1 } {d^3 {\bf{R}}\,} \\
&\quad \left\{ {\cal G}_\omega ^{\left( \eta  \right)} \left( {\left. {\bf{r}} \right|{\bf{R}}} \right) \cdot {\cal G}_\omega ^{\left( \eta  \right)T*} \left( {\left. {{\bf{r}}'} \right|{\bf{R}}} \right) + \frac{{\left[ {{\cal G}_\omega ^{\left( \eta  \right)} \left( {\left. {\bf{r}} \right|{\bf{R}}} \right) \times \mathord{\buildrel{\lower3pt\hbox{$\scriptscriptstyle\leftarrow$}} \over \nabla } _{\bf{R}} } \right] \cdot \left[ {{\cal G}_\omega ^{\left( \eta  \right)} \left( {\left. {{\bf{r}}'} \right|{\bf{R}}} \right) \times \mathord{\buildrel{\lower3pt\hbox{$\scriptscriptstyle\leftarrow$}} \over \nabla } _{\bf{R}} } \right]^{T*} }}{{k_\omega ^2 \left( {1 + \eta ^2 } \right)}} \right\}.
\end{split}
\end{equation}
Although the observation points ${\bf{r}}$ and ${\bf{r}}'$ can be arbitrarily located, any singularities of the dyadic Green's functions at ${\bf{R}} = {\bf{r}}$ and ${\bf{R}} = {\bf{r}}'$ are integrable. Because $D_1$ is a finite spatial volume, the integral necessarily evaluates to a finite quantity, which is strictly nullified by the linear prefactor $\eta$ in the limit $\eta \to 0^+$:
\begin{equation}
\lim_{\eta  \to 0^ +  } \left[ {{\bf{\hat E}}_{\omega s1}^{\left( \eta  \right)} \left( {\bf{r}} \right),{\bf{\hat E}}_{\omega' s1}^{\left( \eta  \right)\dag } \left( {{\bf{r}}'} \right)} \right] = 0,
\end{equation}
demonstrating that the finite region $D_1$ yields no effective contribution in the same limit. For domains $D_2$ and $D_3$, evaluating the commutators using the field expression in Eq.~(\ref{Fic_fiel_23}) yields
\begin{equation}
\left[ {{\bf{\hat E}}_{\omega sj}^{\left( \eta  \right)} \left( {\bf{r}} \right),{\bf{\hat E}}_{\omega 'sj}^{\left( \eta  \right)\dag } \left( {{\bf{r}}'} \right)} \right] = \int {do_{\bf{n}} } \int {do_{{\bf{n}}'} } {\cal F}_{\omega s} \left( {\left. {\bf{r}} \right|{\bf{n}}} \right) \cdot \left[ {{\bf{g}}_{\omega sj}^{\left( \eta  \right)} \left( {\bf{n}} \right),{\bf{g}}_{\omega 'sj}^{\left( \eta  \right)\dag } \left( {{\bf{n}}'} \right)} \right] \cdot {\cal F}_{\omega 's}^{T*} \left( {\left. {{\bf{r}}'} \right|{\bf{n}}'} \right).
\end{equation}
Substituting the operator commutation relation derived in Eq.~(\ref{Comm_g^eta_j}), the angular and frequency Dirac deltas collapse the double integral (${\bf{n}}' = {\bf{n}}$, $\omega' = \omega$), which, together with the transversality condition ${\cal F}_{\omega s} \cdot {\cal I}_{\bf{n}} = {\cal F}_{\omega s}$, simplifies the expression exactly to
\begin{equation}  \label{Comm_E^eta_j}
\left[ {{\bf{\hat E}}_{\omega sj}^{\left( \eta  \right)} \left( {\bf{r}} \right),{\bf{\hat E}}_{\omega 'sj}^{\left( \eta  \right)\dag } \left( {{\bf{r}}'} \right)} \right] = \Lambda_{\omega j} \left( \eta  \right)\delta \left( {\omega  - \omega '} \right)\int {do_{\bf{n}} } {\cal F}_{\omega s} \left( {\left. {\bf{r}} \right|{\bf{n}}} \right) \cdot {\cal F}_{\omega s}^{T*} \left( {\left. {{\bf{r}}'} \right|{\bf{n}}} \right).
\end{equation}
Equation~(\ref{Comm_E^eta_j}) allows us to understand how the sources within the intermediate shell $D_2$ and the purely asymptotic region $D_3$ combine to yield the vacuum fluctuations in Eq.~(\ref{Vac_flu}) in the limit $\eta \to 0^+$, simply by analyzing the factors $\Lambda_{\omega 2}(\eta)$ and $\Lambda_{\omega 3}(\eta)$ defined in Eq.~(\ref{Lamb_j}). Since $\bar{R}_\omega$ is a fixed macroscopic radius, the term $e^{-2\eta k_\omega \bar{R}_\omega}$ strictly converges to unity as $\eta \to 0^+$. Thus, the effective spatial localization of the sources is entirely dictated by the asymptotic behavior of the dynamic exponential $e^{-2\eta k_\omega R_\omega(\eta)}$. It should be emphasized that the specific functional form of the expanding boundary $R_\omega(\eta)$ is entirely arbitrary, provided it satisfies the constraints in Eq.~(\ref{Radii}). Consequently, while we are free to select its precise dependence on $\eta$, the resulting spatial distribution of the sources invariably falls into one of three distinct mathematical classes, governed by the expansion rate:
\begin{enumerate}
    \item \textbf{Fast expansion} ($\lim_{\eta \to 0^+} \eta R_\omega(\eta) = +\infty$): The dynamic exponential vanishes ($e^{-2\eta k_\omega R_\omega(\eta)} \to 0$). Consequently, $\Lambda_{\omega 2} \to 1$ and $\Lambda_{\omega 3} \to 0$. The vacuum fluctuations originating from $D_3$ are completely suppressed, and the entire non-vanishing contribution is localized within the intermediate shell $D_2$.
    \item \textbf{Intermediate expansion} ($\lim_{\eta \to 0^+} \eta R_\omega(\eta) = R_0 > 0$): The dynamic exponential converges to a constant fraction $e^{-2 k_\omega R_0} \in (0,1)$. In this case, $\Lambda_{\omega 2} \to 1 - e^{-2 k_\omega R_0}$ and $\Lambda_{\omega 3} \to e^{-2 k_\omega R_0}$, meaning the effective contribution is split between both domains.
    \item \textbf{Slow expansion} ($\lim_{\eta \to 0^+} \eta R_\omega(\eta) = 0$): The dynamic exponential approaches unity ($e^{-2\eta k_\omega R_\omega(\eta)} \to 1$). Consequently, $\Lambda_{\omega 2} \to 0$ and $\Lambda_{\omega 3} \to 1$. The contribution of the intermediate shell $D_2$ is strictly nullified, and the vacuum fluctuations originate exclusively from the asymptotic region $D_3$.
\end{enumerate}
These results provide a rigorous foundation for determining which sources effectively contribute to the fictitious scattering field ${\bf{\hat E}}_{\omega s}^{\left( \eta  \right)} \left( {\bf{r}} \right)$. For compactness, we assume a power-law functional form 
\begin{equation}
R_\omega\left( \eta  \right) = \left( \eta ^\alpha  k_\omega \right)^{ - 1}, \quad \text{with} \quad \alpha > 0,
\end{equation}
for the radius of the expanding boundary, such that the fast, intermediate, and slow expansion regimes correspond to $\alpha > 1$, $\alpha = 1$, and $\alpha < 1$, respectively. The preceding analysis establishes that, in the limit $\eta \to 0^+$, the fictitious scattering field can be expressed as
\begin{equation}
{\bf{\hat E}}_{\omega s}^{\left( \eta  \right)} \left( {\bf{r}} \right) = \int {do_{\bf{n}} } {\cal F}_{\omega s} \left( {\left. {\bf{r}} \right|{\bf{n}}} \right) \cdot {\bf{g}}_{\omega s}^{\left( \eta  \right)} \left( {\bf{n}} \right),
\end{equation}
where the effective operator ${\bf{g}}_{\omega s}^{\left( \eta  \right)} \left( {\bf{n}} \right)$ is dictated by the expansion parameter $\alpha$ according to
\begin{equation}
{\bf{g}}_{\omega s}^{\left( \eta  \right)} \left( {\bf{n}} \right) = 
\begin{cases}
{\bf{g}}_{\omega s2}^{\left( \eta  \right)} \left( {\bf{n}} \right), & \alpha > 1, \\[6pt]
{\bf{g}}_{\omega s2}^{\left( \eta  \right)} \left( {\bf{n}} \right) + {\bf{g}}_{\omega s3}^{\left( \eta  \right)} \left( {\bf{n}} \right), & \alpha = 1, \\[6pt]
{\bf{g}}_{\omega s3}^{\left( \eta  \right)} \left( {\bf{n}} \right), & \alpha < 1.
\end{cases}
\end{equation}
In all three cases, the limiting behavior of the commutators universally evaluates to
\begin{equation}
\begin{aligned}
 \lim_{\eta \to 0^+} \left[ {{\bf{\hat E}}_{\omega s}^{\left( \eta  \right)} \left( {\bf{r}} \right),{\bf{\hat E}}_{\omega 's}^{\left( \eta  \right)\dag } \left( {{\bf{r}}'} \right)} \right] &= \delta \left( {\omega  - \omega '} \right)\int {do_{\bf{n}} } {\cal F}_{\omega s} \left( {\left. {\bf{r}} \right|{\bf{n}}} \right) \cdot {\cal F}_{\omega s}^{T*} \left( {\left. {{\bf{r}}'} \right|{\bf{n}}} \right), \\ 
 \lim_{\eta \to 0^+} \left[ {{\bf{g}}_{\omega s}^{\left( \eta  \right)} \left( {\bf{n}} \right),{\bf{g}}_{\omega 's}^{\left( \eta  \right)\dag } \left( {{\bf{n}}'} \right)} \right] &= \delta \left( {\omega  - \omega '} \right)\delta \left( {o_{\bf{n}}  - o_{{\bf{n}}'} } \right){\cal I}_{\bf{n}}, 
 \end{aligned}
\end{equation}
where the first relation exactly reproduces the MLNF field commutator $[ {{\bf{\hat E}}_{\omega s} \left( {\bf{r}} \right),{\bf{\hat E}}_{\omega 's}^\dag  \left( {{\bf{r}}'} \right)} ]$, while the second exactly fulfills the necessary condition established in Eq.~(\ref{Weak_Bosonic_Algebra}), thereby guaranteeing the asymptotically bosonic nature of these operators. 

This implies that each of the three mathematical choices ($\alpha > 1$, $\alpha = 1$, and $\alpha < 1$) is physically admissible and, importantly, it can be demonstrated that they are strictly equivalent for the purposes of the asymptotic analysis. The essential reason for this equivalence lies in the fact that, due to the linear prefactor $\eta$, sources localized within any finite spatial region yield a vanishing contribution to the fictitious scattering field in the limit $\eta \to 0^+$. In this work, we deliberately adopt the slow expansion regime for formal clarity: it identifies the effective asymptotic sources without retaining any contribution that vanishes asymptotically, and avoids the appearance of the arbitrary $\bar{R}_\omega$, a parameter introduced solely to conduct the analysis and devoid of any physical significance.

Ultimately, a final remark is in order regarding the initially chosen spatial partition, which enabled the unambiguous identification of the asymptotic sources generating the fictitious scattering field. This partition was introduced to systematically analyze all possible expansion regimes. Evaluating the explicit contribution of each domain provides the mathematical justification to discard non-contributing regions, thereby validating the isolation of the purely asymptotic domain without relying on a priori assumptions.

In conclusion, setting $\alpha = 1/2$ without loss of generality and omitting the domain index $j=3$ for notational simplicity, we have established the following asymptotic relations:
\begin{align}
 {\bf{\hat E}}_{\omega s}^{\left( \eta  \right)} \left( {\bf{r}} \right) &= \int {do_{\bf{n}} } {\cal F}_{\omega s} \left( {\left. {\bf{r}} \right|{\bf{n}}} \right) \cdot {\bf{\hat g}}_{\omega s}^{\left( \eta  \right)} \left( {\bf{n}} \right), \nonumber \\ 
 {\bf{\hat g}}_{\omega s}^{\left( \eta  \right)} \left( {\bf{n}} \right) &= \sqrt {\eta k_\omega  } {\cal I}_{\bf{n}}  \cdot \int_{\left( {\sqrt \eta  k_\omega  } \right)^{ - 1} }^{ + \infty } {dRRe^{ik_\omega  R} } e^{ - \eta k_\omega  R} \left[ {i{\bf{\hat f}}_{\omega e}^{\left( \eta  \right)} \left( { - R{\bf{n}}} \right) - {\bf{n}} \times {\bf{\hat f}}_{\omega m}^{\left( \eta  \right)} \left( { - R{\bf{n}}} \right)} \right], \nonumber \\ 
 \left[ {{\bf{\hat g}}_{\omega s}^{\left( \eta  \right)} \left( {\bf{n}} \right),{\bf{\hat g}}_{\omega 's}^{\left( \eta  \right)\dag } \left( {{\bf{n}}'} \right)} \right] &= \Lambda \left( \eta  \right)\delta \left( {\omega  - \omega '} \right)\delta \left( {o_{\bf{n}}  - o_{{\bf{n}}'} } \right){\cal I}_{\bf{n}}, 
\end{align}
where the finite-$\eta$ prefactor $\Lambda \left( \eta  \right) = e^{ - 2\sqrt \eta  }$ satisfies
\begin{equation}
\mathop {\lim }\limits_{\eta  \to 0^ +  } \Lambda \left( \eta  \right) = 1.
\end{equation}
Note that the factor $\sqrt{(1+i\eta)/(1-i\eta)}$ appearing in Eq.~(\ref{Def_g^eta_j}) has been omitted, as it rigorously converges to unity and yields no contribution in the considered asymptotic limit.

\section{Proof of the asymptotic convergence of the ${\rm LNF}^{(\eta)}$ to the MLNF}
\label{AppD}

In this Appendix, we provide the rigorous mathematical proof that any arbitrary matrix element evaluated within the ${\rm LNF}^{(\eta)}$ framework exactly converges to its MLNF counterpart in the limit $\eta \to 0^+$. The proof relies on introducing a multi-polariton basis within the MLNF Fock space, formally mapping this basis into the ${\rm LNF}^{(\eta)}$ framework, and subsequently evaluating the exact analytical form of generic matrix elements to demonstrate their convergence.

\subsection{Algebraic structure of the MLNF Fock space}
To provide a comprehensive algebraic description of the MLNF Fock space, we define collective phase-space variables and corresponding scalar annihilation operators for the scattering ($s$) and the material ($ e, m$) degrees of freedom. For the scattering sector, characterized by the frequency $\omega$, the propagation direction ${\bf{n}}$, and two transverse polarization states labeled by $\lambda \in \{1, 2\}$ with basis vectors ${\bf{e}}_{{\bf{n}}\lambda}$ (${\bf{n}} \cdot {\bf{e}}_{{\bf{n}}\lambda}=0$), the scalar variables and their associated algebraic structures are defined as follows:
\begin{equation}
\xi^s = \left( \omega, {\bf{n}}, \lambda \right), \quad 
\hat{a}^s \left( \xi^s \right) = {\bf{e}}_{{\bf{n}}\lambda} \cdot {\bf{\hat{g}}}_{\omega s} \left( {\bf{n}} \right), \quad 
\delta^s \left( \xi^s - \xi'^s \right) = \delta \left( \omega - \omega' \right) \delta \left( o_{\bf{n}} - o_{{\bf{n}}'} \right) \delta_{\lambda \lambda'}, \quad 
\int d\xi^s = \int_0^{+\infty} d\omega \int do_{\bf{n}} \sum_{\lambda=1}^2 .
\end{equation}
Here, the first identity introduces $\xi^s$ as a collective coordinate grouping the scattering degrees of freedom. The second equation defines the scalar annihilation operator $\hat{a}^s(\xi^s)$ by projecting the scattering polariton operator ${\bf{\hat{g}}}_{\omega s}({\bf{n}})$ onto the corresponding transverse polarization unit vector. The third relation establishes the generalized Dirac delta distribution over this phase space, factorized into continuous Dirac deltas for the frequency and solid angle, alongside a Kronecker delta for the discrete polarization index. Finally, the fourth equation specifies the total integration measure over the complete scattering configuration space. Similarly, for the material sectors $\nu = e, m$, the degrees of freedom depend on the frequency $\omega$, the spatial coordinate ${\bf{r}}$, and three Cartesian polarization components $\lambda \in \{1, 2, 3\}$ with unit vectors ${\bf{u}}_\lambda$:
\begin{equation}
\xi^\nu = \left( \omega, {\bf{r}}, \lambda \right), \quad \hat{a}^\nu \left( \xi^\nu \right) = {\bf{u}}_\lambda \cdot {\bf{\hat{f}}}_{\omega \nu} \left( {\bf{r}} \right), \quad \delta^\nu \left( \xi^\nu - \xi'^\nu \right) = \delta \left( \omega - \omega' \right) \delta \left( {\bf{r}} - {\bf{r}}' \right) \delta_{\lambda \lambda'}, \quad \int d\xi^\nu = \int_0^{+\infty} d\omega \int d^3{\bf{r}} \sum_{\lambda=1}^3.
\end{equation}
Analogously to the scattering sector, the first expression defines $\xi^\nu$ as the collective phase-space coordinate grouping the material degrees of freedom. The second equation introduces the scalar annihilation operator $\hat{a}^\nu(\xi^\nu)$ by projecting the material polariton operator ${\bf{\hat{f}}}_{\omega \nu}({\bf{r}})$ onto the Cartesian polarization unit vector. The third relation establishes the generalized Dirac delta distribution for the material phase space, which factorizes into continuous Dirac deltas for the frequency and three-dimensional spatial coordinates, along with a Kronecker delta for the discrete Cartesian index. Finally, the fourth equation specifies the total integration measure over the material configuration space, comprising continuous integrations over the positive frequency domain and the full tridimensional volume, completed by a discrete summation over the three spatial polarization components.
Using a unified sector index $\mu \in \{s, e, m\}$, the canonical bosonic commutation relations for the exact MLNF operators take the compact universal form:
\begin{equation} \label{Uni_Com}
\left[ \hat{a}^\mu \left( \xi^\mu \right), \hat{a}^{\mu'\dag} \left( \xi'^{\mu'} \right) \right] = \delta^{\mu\mu'} \delta^\mu \left( \xi^\mu - \xi'^{\mu'} \right).
\end{equation}
We now construct the basis states for the complete multi-sector MLNF Fock space. A generic state containing $p$ scattering polaritons, $q$ electric polaritons, and $r$ magnetic polaritons is parameterized by the collective configuration coordinate $\Xi_{pqr}$, representing a specific set of polariton variables,
\begin{equation}
\Xi_{pqr} = \left( \xi_1^s, \ldots, \xi_p^s, \xi_1^e, \ldots, \xi_q^e, \xi_1^m, \ldots, \xi_r^m \right).
\end{equation}
Such a basis state is generated by the sequential application of the scalar creation operators onto the MLNF vacuum $|0\rangle$:
\begin{equation}
\left| \Xi_{pqr} \right\rangle = \left[ \frac{1}{\sqrt{p!}} \prod_{i=1}^p \hat{a}^{s\dag} \left( \xi_i^s \right) \right] \left[ \frac{1}{\sqrt{q!}} \prod_{j=1}^q \hat{a}^{e\dag} \left( \xi_j^e \right) \right] \left[ \frac{1}{\sqrt{r!}} \prod_{k=1}^r \hat{a}^{m\dag} \left( \xi_k^m \right) \right] \left| 0 \right\rangle.
\end{equation}
By employing the canonical commutation relations in Eq.~(\ref{Uni_Com}), the inner product between two such basis states decomposes into independent sectorial contributions, each governed by a symmetrized Dirac delta distribution over all particle permutations $\pi$ within the respective symmetric groups $S_p, S_q, S_r$:
\begin{equation}
\left\langle \Xi_{pqr} \middle| \Xi'_{p'q'r'} \right\rangle = 
\left[ \frac{\delta_{p,p'}}{p!} \sum_{\pi \in S_p} \prod_{i=1}^p \delta^s \left( \xi_i^s - \xi'^s_{\pi(i)} \right) \right] 
\left[ \frac{\delta_{q,q'}}{q!} \sum_{\pi \in S_q} \prod_{j=1}^q \delta^e \left( \xi_j^e - \xi'^e_{\pi(j)} \right) \right] 
\left[ \frac{\delta_{r,r'}}{r!} \sum_{\pi \in S_r} \prod_{k=1}^r \delta^m \left( \xi_k^m - \xi'^m_{\pi(k)} \right) \right].
\end{equation}
This expression precisely encodes the orthonormality condition in the continuous macroscopic bosonic sense, ensuring that states are strictly orthogonal unless the number of excitations matches exactly within each independent sector (i.e., $p=p'$, $q=q'$, and $r=r'$); when these particle numbers coincide, the inner product is totally symmetric under the exchange of any identical polaritonic variables. These orthogonal basis states satisfy the completeness relation, which resolves the identity operator $\hat{I}$ over the full Fock space:
\begin{equation} \label{complet}
\sum_{p,q,r=0}^\infty \int d\Xi_{pqr} \left| \Xi_{pqr} \right\rangle \! \left\langle \Xi_{pqr} \right| = \hat{I},
\end{equation}
where the total integration measure implies independent integrations over each subset of polaritonic variables:
\begin{equation}
\int d\Xi_{pqr} = \left[ \prod_{i=1}^p \int d\xi_i^s \right] \left[ \prod_{j=1}^q \int d\xi_j^e \right] \left[ \prod_{k=1}^r \int d\xi_k^m \right].
\end{equation}
Any arbitrary physical state $|\Psi\rangle$ can thus be expanded in this basis via its multi-particle probability amplitudes $\Psi_{pqr}(\Xi_{pqr})$:
\begin{equation}
\left| \Psi \right\rangle = \sum_{p,q,r=0}^\infty \int d\Xi_{pqr} \; \Psi_{pqr} \left( \Xi_{pqr} \right) \left| \Xi_{pqr} \right\rangle,
\end{equation}
where, by construction, the amplitudes $\Psi_{pqr}(\Xi_{pqr})$ are totally symmetric under the permutation of any identical polaritonic variables within each of the three sectors. This state is subject to the standard normalization condition:
\begin{equation}
\sum_{p,q,r=0}^\infty \int d\Xi_{pqr} \; \left| \Psi_{pqr} \left( \Xi_{pqr} \right) \right|^2 = 1.
\end{equation}
Similarly, an arbitrary observable $\hat{O}$ is defined by its representation in the chosen continuous basis, $O(\Xi_{pqr}, \Xi'_{p'q'r'})$:
\begin{equation}
\hat{O} = \sum_{p,q,r=0}^\infty \sum_{p',q',r'=0}^\infty \int d\Xi_{pqr} \int d\Xi'_{p'q'r'} O \left( \Xi_{pqr}, \Xi'_{p'q'r'} \right) \left| \Xi_{pqr} \right\rangle \left\langle \Xi'_{p'q'r'} \right|.
\end{equation}
Consequently, the MLNF matrix element between two arbitrary states $|\Phi\rangle$ and $|\Psi\rangle$ is rigorously given by:
\begin{equation} \label{MLNF_mat_ele}
\left\langle \Phi \right| \hat{O} \left| \Psi \right\rangle = \sum_{p,q,r=0}^\infty \sum_{p',q',r'=0}^\infty \int d\Xi_{pqr} \int d\Xi'_{p'q'r'} \Phi_{pqr}^* \left( \Xi_{pqr} \right)  O \left( \Xi_{pqr}, \Xi'_{p'q'r'} \right) \Psi_{p'q'r'} \left( \Xi'_{p'q'r'} \right).
\end{equation}

\subsection{MLNF mapping into the ${\rm LNF}^{(\eta)}$ framework}
Having established the algebraic structure of the MLNF, we now introduce its formal mapping to the ${\rm LNF}^{(\eta)}$ framework. Since the entire MLNF Fock space and the associated operator algebra are exhaustively generated by the fundamental polariton operators acting on the vacuum, it is mathematically sufficient to define the mapping solely by specifying the images of these generators. Accordingly, the continuous polariton operators and the vacuum state are mapped to their ${\rm LNF}^{(\eta)}$ counterparts as follows:
\begin{equation}
{\bf{\hat{g}}}_{\omega s} \left( {\bf{n}} \right) \mapsto {\bf{\hat{g}}}_{\omega s}^{(\eta)} \left( {\bf{n}} \right), \quad \quad \quad
{\bf{\hat{f}}}_{\omega \nu} \left( {\bf{r}} \right) \mapsto {\bf{\hat{f}}}_{\omega \nu}^{(\eta)} \left( {\bf{r}} \right), \quad \quad \quad
\left| 0 \right\rangle \mapsto \left| 0^{(\eta)} \right\rangle,
\end{equation}
where the image operators ${\bf{\hat{g}}}_{\omega s}^{(\eta)}$ are precisely the asymptotic scattering operators derived in Sec.~IV [see Eq.~(\ref{g^(eta)})], while ${\bf{\hat{f}}}_{\omega \nu}^{(\eta)}$ are the standard material polariton operators natively defined within the ${\rm LNF}^{(\eta)}$ formulation. Consequently, the mapping for the scalar polariton operators naturally follows as:
\begin{equation} 
\hat{a}^s \left( \xi^s \right) \mapsto \hat{a}^{s(\eta)} \left( \xi^s \right) = {\bf{e}}_{{\bf{n}}\lambda} \cdot {\bf{\hat{g}}}_{\omega s}^{(\eta)} \left( {\bf{n}} \right), \quad \quad \quad \hat{a}^\nu \left( \xi^\nu \right) \mapsto \hat{a}^{\nu(\eta)} \left( \xi^\nu \right) = {\bf{u}}_\lambda \cdot {\bf{\hat{f}}}_{\omega \nu}^{(\eta)} \left( {\bf{r}} \right).
\end{equation}
As a direct consequence of the commutation relations in Eqs.~(\ref{Comm_g^(eta)}) and (\ref{Comm_f^(eta)}), the mapped scalar scattering operators are not strictly bosonic for finite $\eta$ due to the acquired structural factor $\Lambda(\eta)$, whereas the scalar material polariton operators strictly preserve their exact bosonic nature. Using again the unified sector index $\mu \in \{s, e, m\}$, the commutation relations for mapped scalar polariton operators take the form:
\begin{equation}
\left[ \hat{a}^{\mu(\eta)} \left( \xi^\mu \right), \hat{a}^{\mu'(\eta)\dag} \left( \xi'^{\mu'} \right) \right] = \delta^{\mu \mu'} \left[ \delta^{\mu s} \Lambda(\eta) + \left(1 - \delta^{\mu s} \right) \right] \delta^\mu \left( \xi^\mu - \xi'^{\mu'} \right).
\end{equation}
However, it is crucial to emphasize that the scalar scattering operators are asymptotically bosonic, since the relation $\lim_{\eta \to 0^+} \Lambda(\eta) = 1$ naturally restores the exact bosonic commutation relations. Proceeding further, the basis states of the MLNF Fock space are consequently mapped to the corresponding states in the ${\rm LNF}^{(\eta)}$ framework:
\begin{equation}
\left| \Xi_{pqr} \right\rangle \mapsto \left| \Xi_{pqr}^{(\eta)} \right\rangle = \left[ \frac{1}{\sqrt{p!}} \prod_{i=1}^p \hat{a}^{s(\eta)\dag} \left( \xi_i^s \right) \right] \left[ \frac{1}{\sqrt{q!}} \prod_{j=1}^q \hat{a}^{e(\eta)\dag} \left( \xi_j^e \right) \right] \left[ \frac{1}{\sqrt{r!}} \prod_{k=1}^r \hat{a}^{m(\eta)\dag} \left( \xi_k^m \right) \right] \left| 0^{(\eta)} \right\rangle.
\end{equation}
When evaluating the inner product between two mapped basis states, namely $\langle \Xi_{pqr}^{(\eta)} | \Xi'^{(\eta)}_{p'q'r'} \rangle$, the creation and annihilation operators must be contracted down to the vacuum. Because the material operators ($\nu = e, m$) preserve their exact bosonic algebra and commute with the scattering sector, their contractions yield the standard symmetrized combinations of Dirac deltas. Conversely, within the scattering sector, although the mapped operators recover the standard bosonic algebra only in the limit $\eta \to 0^+$, their finite-$\eta$ commutator evaluates to a scalar distribution strictly proportional to the identity. This property ensures that the algebraic structure underpinning Wick's theorem is fully preserved even for strictly positive $\eta$. Therefore, every contraction between a mapped annihilation and creation operator is governed by the mapped commutator $\left[ \hat{a}^{s(\eta)}, \hat{a}^{s(\eta)\dag} \right]$, extracting an explicit structural factor $\Lambda(\eta)$. Since a non-vanishing overlap strictly requires the number of scattering quanta in the bra and the ket to match ($p=p'$), the complete contraction of the scattering operators systematically accumulates exactly $p$ factors of $\Lambda(\eta)$. For instance, considering the overlap of two generic normalized states of $p=p'=2$ scattering quanta, the application of Wick's theorem explicitly yields:
\begin{align}
&\left\langle {0^{\left( \eta  \right)} } \right|\frac{1}{{\sqrt {2!} }}\hat a^{s\left( \eta  \right)} \left( {\xi _1^s } \right)\hat a^{s\left( \eta  \right)} \left( {\xi _2^s } \right)\frac{1}{{\sqrt {2!} }}\hat a^{s\left( \eta  \right)\dag } \left( {{\xi '}_1^s } \right)\hat a^{s\left( \eta  \right)\dag } \left( {{\xi '}_2^s } \right)\left| {0^{\left( \eta  \right)} } \right\rangle \nonumber \\ 
&\quad = \frac{1}{2!} \left( \left[ {\hat a^{s\left( \eta  \right)} \left( {\xi _1^s } \right),\hat a^{s\left( \eta  \right)\dag } \left( {{\xi '}_1^s } \right)} \right]\left[ {\hat a^{s\left( \eta  \right)} \left( {\xi _2^s } \right),\hat a^{s\left( \eta  \right)\dag } \left( {{\xi '}_2^s } \right)} \right] + \left[ {\hat a^{s\left( \eta  \right)} \left( {\xi _1^s } \right),\hat a^{s\left( \eta  \right)\dag } \left( {{\xi '}_2^s } \right)} \right]\left[ {\hat a^{s\left( \eta  \right)} \left( {\xi _2^s } \right),\hat a^{s\left( \eta  \right)\dag } \left( {{\xi '}_1^s } \right)} \right] \right) \nonumber \\ 
&\quad = \Lambda ^2 \left( \eta  \right) \frac{1}{2!}  \left[ {\delta \left( {\xi _1^s  - {\xi '}_1^s } \right)\delta \left( {\xi _2^s  - {\xi '}_2^s } \right) + \delta \left( {\xi _1^s  - {\xi '}_2^s } \right)\delta \left( {\xi _2^s  - {\xi '}_1^s } \right)} \right] \nonumber \\ 
&\quad = \Lambda ^2 \left( \eta  \right) \left\langle 0 \right|\frac{1}{{\sqrt {2!} }}\hat a\left( {\xi _1^s } \right)\hat a\left( {\xi _2^s } \right)\frac{1}{{\sqrt {2!} }}\hat a^\dag  \left( {{\xi '}_1^s } \right)\hat a^\dag  \left( {{\xi '}_2^s } \right)\left| 0 \right\rangle,
\end{align}
manifestly demonstrating how each fully contracted permutation is homogeneously scaled by the overall macroscopic factor $\Lambda^2(\eta)$. As a result, the inner product of the mapped basis states factorizes perfectly, reducing to the exact canonical MLNF inner product scaled by the overall factor $[\Lambda(\eta)]^p$:
\begin{equation} \label{fund_sca_map}
\left\langle \Xi_{pqr}^{(\eta)} \middle| \Xi'^{(\eta)}_{p'q'r'} \right\rangle = \left[ \Lambda(\eta) \right]^p \left\langle \Xi_{pqr} \middle| \Xi'_{p'q'r'} \right\rangle,
\end{equation}
which is a central mathematical result of our analysis. Finally, arbitrary states and operators within the MLNF are mapped into the ${\rm LNF}^{(\eta)}$ framework by exploiting their linear expansion over the complete Fock space basis. Explicitly, a general state $\left| \Psi \right\rangle$ and a general operator $\hat{O}$ are formally mapped as follows:
\begin{equation}
\begin{split}
\left| \Psi \right\rangle &\mapsto \left| \Psi^{(\eta)} \right\rangle = \sum_{p,q,r = 0}^\infty \int d\Xi_{pqr} \; \Psi_{pqr} \left( \Xi_{pqr} \right) \left| \Xi_{pqr}^{(\eta)} \right\rangle, \\
\hat{O} &\mapsto \hat{O}^{(\eta)} = \sum_{p,q,r = 0}^\infty \sum_{p',q',r' = 0}^\infty \int d\Xi_{pqr} \int d\Xi'_{p'q'r'} \; O \left( \Xi_{pqr}, \Xi'_{p'q'r'} \right) \left| \Xi_{pqr}^{(\eta)} \right\rangle \left\langle \Xi'^{(\eta)}_{p'q'r'} \right|.
\end{split}
\end{equation}
It is crucial to note that this mapping exclusively affects the basis kets and bras, leaving the scalar expansion coefficients—namely, the probability amplitudes $\Psi_{pqr} \left( \Xi_{pqr} \right)$ and the matrix elements $O \left( \Xi_{pqr}, \Xi'_{p'q'r'} \right)$—strictly unaltered. These coefficients encode the specific physical configurations of the state and the observable, respectively, and are therefore intrinsically invariant under the formal mapping from the MLNF into the ${\rm LNF}^{(\eta)}$ framework.

\subsection{Convergence of matrix elements}
Let us now consider the mapped matrix element $\langle \Phi^{(\eta)} | \hat{O}^{(\eta)} | \Psi^{(\eta)} \rangle$ for arbitrary physical observables, constructed via the mapped states $|\Phi^{(\eta)}\rangle, |\Psi^{(\eta)}\rangle$ and the mapped operator $\hat{O}^{(\eta)}$. By inserting the completeness relations (see Eq.~(\ref{complet})) into the matrix element and applying the fundamental mapping identity of Eq.~(\ref{fund_sca_map}) to evaluate the inner products of the mapped basis states, the structural factor $[\Lambda(\eta)]^{p}$ naturally emerges from the left inner product $\langle \Phi^{(\eta)} | \Xi^{(\eta)} \rangle$, while the factor $[\Lambda(\eta)]^{p'}$ emerges from the right inner product $\langle \Xi'^{(\eta)} | \Psi^{(\eta)} \rangle$. This yields the exact analytical form of the generic mapped expectation value:
\begin{equation}
\left\langle \Phi^{(\eta)} \right| \hat{O}^{(\eta)} \left| \Psi^{(\eta)} \right\rangle = \sum_{p,q,r=0}^\infty \sum_{p',q',r'=0}^\infty \left[ \Lambda(\eta) \right]^{p+p'} \int d\Xi_{pqr} \int d\Xi'_{p'q'r'} \Phi_{pqr}^* \left( \Xi_{pqr} \right) O \left( \Xi_{pqr}, \Xi'_{p'q'r'} \right)  \Psi_{p'q'r'} \left( \Xi'_{p'q'r'} \right).
\end{equation}
Evaluating the asymptotic limit $\eta \to 0^+$ identically yields $\left[ \Lambda(\eta) \right]^{p+p'} \to 1$ for all subspace sectors, thereby retrieving the exact MLNF matrix element defined in Eq.~(\ref{MLNF_mat_ele}). This evaluation is mathematically rigorous because the limit can be safely interchanged with the summations over the particle number indices; such an interchange is guaranteed by the fact that the infinite series defining the MLNF matrix element converges to a finite physical value. Consequently, the mapped expression asymptotically reduces to its original counterpart:
\begin{equation}
\lim_{\eta \to 0^+} \left\langle \Phi^{(\eta)} \right| \hat{O}^{(\eta)} \left| \Psi^{(\eta)} \right\rangle = \left\langle \Phi \right| \hat{O} \left| \Psi \right\rangle.
\end{equation}
This constitutes the rigorous proof that any matrix element evaluated within the ${\rm LNF}^{(\eta)}$ framework identically converges to the corresponding counterpart of the MLNF theory in the asymptotic limit.

\end{document}